\documentclass[journal]{IEEEtran}
\usepackage{amsmath,amsfonts}
\usepackage{algorithmic}
\usepackage{array}
\usepackage{stfloats}
\usepackage{url}
\usepackage{verbatim}
\usepackage{balance}

\usepackage{cancel}

\newcommand{\dd}{\textnormal{d}}

\usepackage{amssymb} 
\usepackage[center]{caption} 
\usepackage{tikz}
\usetikzlibrary{arrows.meta}
\allowdisplaybreaks
\usepackage[amsmath]{ntheorem}
\usepackage{dsfont}
\usepackage{bbm}
\usepackage{mleftright}
\mleftright

\usepackage{hyperref}
\usepackage{cleveref}
\hypersetup{
    colorlinks=true, 
    linktoc=all,     
    linkcolor=black,  
    citecolor=black
}

\theoremstyle{plain}
\theoremheaderfont{\normalfont\itshape}
\theoremseparator{:}
\newtheorem{theorem}{Theorem}
\newtheorem{lemma}{Lemma}
\newtheorem{definition}{Definition}
\newtheorem{remark}{Remark}
\newtheorem{assumption}{Assumption}
\newtheorem{proposition}{Proposition}

\newcommand{\sizecorr}[1]{\makebox[0cm]{\phantom{$\displaystyle #1$}}}

\def\BibTeX{{\rm B\kern-.05em{\sc i\kern-.025em b}\kern-.08em
    T\kern-.1667em\lower.7ex\hbox{E}\kern-.125emX}}

\begin{document}

\title{Covert Communication Over Additive-Noise Channels}

\author{C\'ecile~Bouette,
        Laura~Luzzi,
        and~Ligong~Wang
\thanks{C\'ecile Bouette and Laura Luzzi are with Laboratoire ETIS---UMR 8051, CY Cergy Paris Universit\'e, ENSEA, CNRS, 95000 Cergy-Pontoise, France (e-mail: \mbox{cecile.bouette@ensea.fr}; \mbox{laura.luzzi@ensea.fr}). Laura Luzzi is currently a visiting researcher with Project COSMIQ, Centre Inria de Paris, 48 rue Barrault, 75013 Paris, France (e-mail: \mbox{laura.luzzi@inria.fr}).}
\thanks{Ligong Wang is with the Department of Information Technology and Electrical Engineering, ETH Zurich, 8092 Zurich, Switzerland (e-mail: \mbox{ligwang@isi.ee.ethz.ch}).}
\thanks{This work was supported in part by CY Initiative of Excellence (grant Investissements d’Avenir ANR-16-IDEX-0008).}}

\markboth{To appear in IEEE Transactions on Information Theory}%
{Bouette \MakeLowercase{\textit{et al.}}: Covert Communication Over Additive-Noise Channels}

\maketitle

\begin{abstract}
We study the fundamental limits of covert communications over general memoryless additive-noise channels. We assume that the legitimate receiver and the eavesdropper share the same channel and therefore see the same outputs. Under mild integrability assumptions, we find a general upper bound on the square-root scaling constant, which only involves the variance of the logarithm of the probability density function of the noise. Furthermore, we show that, under some additional assumptions, this upper bound is tight. We also provide upper bounds on the length of the secret key required to achieve the optimal scaling. 
\end{abstract}


\section{Introduction}
\IEEEPARstart{C}{overt communication}, also known as communication with low probability of detection \cite{bash_article,che2013reliable,fundamental_covertness,bloch_resolvability}, refers to a scenario where a transmitter and a receiver
wish to prevent a potential eavesdropper from making a good guess on whether communication is ongoing or not. Covertness is desirable in 
many applications, since merely revealing \emph{who} is communicating, \emph{when}, and \emph{from where} can leak sensitive information, even if the content of the communication is not disclosed.  

Covert communication has been previously studied in the context of steganography and spread-spectrum techniques. 
In the framework of information theory, \cite{bash_article} first determined the order of magnitude of the achievable message length that the legitimate users can reliably communicate over a noisy channel while ensuring a low probability of detection by the eavesdropper.
The work showed that the capacity---in nats per channel use---under a covertness constraint is zero: the maximum message length that can be transmitted reliably and covertly scales like the square root of the total number of channel uses. This phenomenon is sometimes called the \textit{square-root law} and is also observed in steganography \cite{steganography_square_root_law,delp_iii_square_2009}. 
While \cite{bash_article} considered additive white Gaussian noise (AWGN) channels, the square-root law was also established for binary symmetric channels in \cite{che2013reliable}, which, in addition, shows that covert communication can sometimes be realized without employing a secret key.
The exact scaling constant for the square-root law---which we shall formally define later on---was characterized for discrete memoryless channels (DMCs) and AWGN channels in \cite{fundamental_covertness}. The minimum required key length for covert communication over various types of channels was derived in \cite{bloch_resolvability}.

Some other works studied situations where the square-root law may not hold, e.g., where the eavesdropper has uncertainty about the noise distribution \cite{article_uncertainty_2}, or in some multiple-antenna scenarios
\cite{MIMO_covertness}. The present work does not consider such scenarios and stays within the framework of \cite{bash_article,che2013reliable,fundamental_covertness,bloch_resolvability}.

In a recent work \cite{bouette2023covert}, we extended the previous results on AWGN channels to channels with \emph{generalized Gaussian} noise to compute or, for some parameter range, upper-bound the square-root scaling constant for covert communication. In the present paper, we make yet a further extension to consider general noise distributions.

We note that non-Gaussian noise models have practical relevance. For instance, the Laplace distribution is used to model noise spikes due to rare events  \cite[Chapter 10]{fundamental_signal}; generalized Gaussian distributions \cite{nadarajah2005,generalized_gaussian_distributions} can be used to model impulsive noise, such as atmospheric noise  \cite{use_of_gg_distribution_atmospheric_noise} and interference in ultra-wideband systems \cite{use_of_gg_distribution_interferences};  dense wireless networks with interference are characterized by heavy-tailed noise, which can be modeled using $\alpha$-stable distributions \cite{cardieri2010modeling, de2017capacity, clavier2021experimental}.
 
In this work, we assume that the legitimate receiver and the eavesdropper face exactly the same noise distribution.
Under mild integrability assumptions, we show that the square-root scaling constant is upper-bounded by a simple expression that only involves the variance of the logarithm of the probability density function (PDF) of the noise. We then show that, under some additional assumptions on the noise PDF, the said upper bound is tight, i.e., there exist covert channel codes that can asymptotically achieve it. We further provide (sometimes loose) upper bounds on the key length that is needed to achieve the optimal scaling constant.

As examples, we compute the scaling constant (or its upper bound) when the noise distribution is exponential, generalized Gaussian, or gamma. For generalized Gaussian noise, we recover the results in \cite{bouette2023covert}.

The rest of this paper is organized as follows.  Section \ref{section:setup} 
introduces the problem setup. Section \ref{section:upper_bound_L} proves an upper bound on the square-root scaling constant for covert communication over general additive-noise channels. Section \ref{section:lower_bound_L} shows that, under some additional assumptions on the noise distribution, the upper bound of Section~\ref{section:upper_bound_L} is achievable. Section \ref{section:examples} computes the scaling constant (or the upper bound) for a few examples. Section \ref{section:key_length} contains upper bounds on the key length required for the achievability result in Section~\ref{section:lower_bound_L}.
Finally, Section \ref{section:conclusion} concludes the paper with some remarks.

\section{Problem Setup}\label{section:setup}
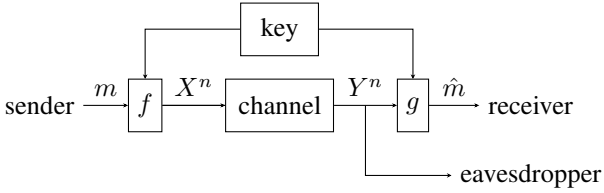
\begin{figure}[tbp]
\begin{center}
            \begin{tikzpicture}[
            nodetype1/.style={
                rectangle,
                minimum width=0.5cm,
                minimum height=0.7cm,
                draw=black,
                font=\normalsize
            },
            nodetype2/.style={
                rectangle,
                minimum width=0.4cm,
                minimum height=0.7cm,
                draw=black,
                font=\normalsize
            },
            tip2/.style={-{Stealth[length=0.6mm, width=0.5mm]}}
            ]
            \matrix[row sep=0.3cm, column sep=0.3cm, ampersand replacement=\&]{
            \& \& (invisible) \&\&
            \node (Key) [draw, nodetype1, text width=0.8cm, text centered]  {\text{key}}; \&\& \\
            \node (Alice) {\text{sender}};  \& \& \node (encoder) [nodetype2]   {$f$}; \&
            \node (X){}; \&
            \node (W) [draw, nodetype1, text width=1.2cm, text centered]  {\text{channel}}; \&
            \node (Y){};
            \&
            \node (decoder) [nodetype2] {$g$}; \&
            \node (Bob) {\text{receiver}};\\
            \& \& \&
            \& \& \& \&
            \node (Eve) {\text{eavesdropper}}; \& \\};
            
            \draw[->] (Alice) edge[tip2] node [above] {$m$} (encoder) ;
            \draw[->] (encoder) edge[tip2] node [above] (X) {} (W) ;
            \draw[-]  (encoder.east) -- node [above] (X1) {$X^n$} (W.west) ;
            \draw[arrows = {-Latex[length=1pt]}] (W) edge[tip2] node [above] {$Y^n$} (decoder) ;
            \draw[->] (decoder) edge[tip2] node [above] {$\hat{m}$} (Bob) ;
            \draw[-{Stealth[length=0.5mm, width=0.5mm]}] (Y.center)  |- node [above] {} (Eve) ;
            \draw[{Stealth[length=0.5mm, width=0.5mm]}-] (encoder) |-  (Key) ;
            \draw[-{Stealth[length=0.5mm, width=0.5mm]}] (Key) -| (decoder) ;
            \end{tikzpicture}
    \caption{General setup for covert communication.}
     \label{fig:general_channel}
     \end{center}
\end{figure}

We usually use upper-case letters like $X$ to denote (real) random variables and lower-case letters like $x$ to denote their realizations. A length-$n$ random vector $(X_1,\ldots,X_n)$ is denoted $X^n$. We use $P_X$ to denote the cumulative distribution function of the random variable $X$ and $P_{X^n}$ that of the random vector $X^n$. The PDF corresponding to $P_X$, when it exists, is denoted $p_X$. The Kullback-Leibler divergence relative entropy) \cite{coverthomas06} between two distributions $P$ and $Q$ is denoted ${D}(P\|Q)$, the differential entropy of random variable $X$ is denoted $h(X)$, and the mutual information between $X$ and $Y$ is denoted $I(X;Y)$; all of these are measured in nats. 

We consider the setup illustrated in Fig.~\ref{fig:general_channel}, where a transmitter and a receiver communicate in the presence of an eavesdropper
over an additive-noise channel described by

\begin{equation}
\label{eq:channel}
Y_i=X_i+Z_i,\qquad i=1,2,\ldots, n,
\end{equation}
where $X_i$ denotes the channel input random variable, $Y_i$ the channel output  random variable, and $Z_i$ the additive noise  random variable, at time $i$, all of which take values in $\mathbb{R}$. 

The sender and the receiver are assumed to share a sufficiently long secret key $\mathsf{K} \in \mathcal{K}$. 
The key and the message are assumed to be uniformly distributed and independent of each other.  
A 
code $\mathcal{C}=(f,g)$ of length $n$ for message set $\mathcal{M}$ and key set $\mathcal{K}$ consists of an encoder $f\colon\mathcal{M} \times \mathcal{K} \rightarrow \mathbb{R}^n , (m,k) \mapsto x^n$ and a decoder $g\colon \mathbb{R}^n \times \mathcal{K} \rightarrow \mathcal{M}, (y^n,k) \mapsto \hat{m}$.

The eavesdropper observes the same output as the legitimate receiver. 
We will assume that the eavesdropper knows the encoding and decoding functions $f$ and $g$, but not the value of the secret key $\mathsf{K}$.

Covertness requires that the eavesdropper should not be able to detect whether transmission is ongoing or not. Specifically, we consider the following covertness condition: 
for some given $\Delta >0$ and for every $n$, the output distribution must satisfy
\begin{equation}
 \label{covert_communication_hypothesis}
     {D}(P_{Y^n}\|P_{Z^n}) \leq \Delta,  
\end{equation}
where $P_{Z^n}$ denotes the distribution of the noise vector $Z^n$, and $P_{Y^n}$ that of the output sequence averaged over the messages and over the key: for every $y^n\in\mathbb{R}^n$,
\begin{IEEEeqnarray}{rCl}
    p_{Y^n}(y^n)=\frac{1}{|\mathcal{K}|\cdot|\mathcal{M}|}\sum_{m=1}^{|\mathcal{M}|}
    \sum_{k=1}^{|\mathcal{K}|}
    p_{Y^n|X^n} (y^n|f(m,k)). \IEEEeqnarraynumspace
\end{IEEEeqnarray}

Given $\epsilon>0$, we denote by $A_n(\Delta, \epsilon)$ the maximum of $\ln|\mathcal{M}|$ for which there exists a code $\mathcal{C}$ of length $n$ that satisfies covertness condition \eqref{covert_communication_hypothesis}, and whose average probability of decoding error is at most $\epsilon$.
As in \cite{fundamental_covertness}, we define the square-root scaling constant:
\begin{equation}
\label{L_definition}
L\triangleq \lim\limits_{\epsilon\downarrow  0} \liminf\limits_{n\rightarrow  \infty} \dfrac{A_n(\Delta,\epsilon)}{\sqrt{n\Delta}}  .
\end{equation}

We assume that the noise stochastic process $\{Z_i\}_{i\in\{1,2,\dots,n\}}$
is independent of the message and the secret key. We further assume that $\{Z_i\}_{i\in\{1,2,\dots,n\}}$ is independent and identically distributed (i.i.d.) according to a PDF $p_Z(z)$, $z\in\mathbb{R}$. A fortiori, $p_Z$ is Lebesgue-measurable and $\int_{\mathbb{R}} p_Z(z) \dd z = 1$.

We make the following technical assumptions on $p_Z$: there exists some $\zeta\in(0,1)$ such that
\begin{IEEEeqnarray}{rCl}
    \label{eq:integrable0}
  \int_{\mathbb{R}} p_Z(z) \left(\ln(p_Z(z))\right)^4 \dd z & < & \infty\\
  \int_{\mathbb{R}} p_Z(z)^\zeta \dd z & < & \infty\label{eq:integrable1}\\
  \int_{\mathbb{R}} p_Z(z)^\zeta \left(\ln(p_Z(z))\right)^4\dd z & < & \infty. \label{eq:integrable2}
  \end{IEEEeqnarray}
These imply some further integrability properties:
\begin{lemma}
\label{lemma_integrability}
If $p_Z$ satisfies \eqref{eq:integrable0}--\eqref{eq:integrable2}, then, for every $k\in\{0,1,2,3,4\}$, the integrals
\begin{IEEEeqnarray}{c}
\int_{\mathbb{R}} p_Z(z)^{\nu(z)} \bigl|\ln (p_Z(z))\bigr|^k\dd z  
\end{IEEEeqnarray}
are uniformly bounded over $\nu\colon \mathbb{R}\to[\zeta,1]$, $z\mapsto \nu(z)$.
\end{lemma}
\begin{IEEEproof}
For any $k\in\{0,1,2,3,4\}$, $\nu\colon \mathbb{R}\to[\zeta,1]$, and $z\in\mathbb{R}$,
\begin{IEEEeqnarray}{rCl}
\IEEEeqnarraymulticol{3}{l}{
p_Z(z)^{\nu(z)} \bigl|\ln(p_Z(z))\bigr|^k
}\nonumber\\* ~~~~~
 &\leq& \left(p_Z(z)+p_Z(z)^\zeta \right)\left(1+(\ln(p_Z(z)))^4\right),
\end{IEEEeqnarray}
where the right-hand side does not depend on $\nu$ and is integrable due to \eqref{eq:integrable0}--\eqref{eq:integrable2}.
 \end{IEEEproof}

\section{A General Upper Bound on $L$}
\label{section:upper_bound_L}

The following theorem provides a general upper bound on~$L$. We shall later show that, under some additional assumptions on $p_Z$, this bound is tight.

\begin{theorem}\label{theorem_converse}
For the memoryless additive-noise channel \eqref{eq:channel} with $p_Z$ satisfying \eqref{eq:integrable0}--\eqref{eq:integrable2},
\begin{equation}\label{eq:upper}
    L \leq \sqrt{2} \sqrt{ \textnormal{Var} \left[\ln(p_Z(Z))\right]}.
\end{equation}
\end{theorem}

The following lemma will be used in the proof of Theorem~\ref{theorem_converse}.

\begin{lemma}
\label{lemma_inequalitiy_entropy_generalized}
Consider $p_Z$ satisfying \eqref{eq:integrable0}--\eqref{eq:integrable2}. For any $\gamma\in[0,1-\zeta)$, let the random variable $\tilde{Z}$ have PDF
\begin{equation}
\label{def_z_tilde}
    p_{\tilde{Z}}(\tilde{z}) =\alpha \cdot p_Z(\tilde{z})^{1-\gamma}, \quad \tilde{z} \in \mathbb{R}, 
\end{equation}
where
\begin{equation}
\label{def_alpha}
\alpha = \left(\int_{\mathbb{R}}p_Z(z)^{1-\gamma} \dd z\right)^{-1}.
\end{equation}
Then the following hold: 
\begin{enumerate}
\item \label{it:lem1} For $\gamma\in[0,1-\zeta)$,
\begin{equation}
\label{entropy_z_tilde}
    h(\tilde{Z}) = -\frac{\ln (\alpha)}{\gamma} + \frac{1-\gamma}{\gamma} \,{D}(P_{\tilde{Z}}\| P_Z).
\end{equation}
\item \label{it:lem2} For any random variable $Y$ satisfying
\begin{equation}\label{eq:Deps}
    {D}(P_Y\| P_Z) \leq {D}(P_{\tilde{Z}}\| P_Z),
\end{equation}
we have
\begin{equation}
\label{entropy_upper_bound}
    h(Y)\leq h(\tilde{Z}).
\end{equation}
That is, $p_{\tilde{Z}}$ as in \eqref{def_z_tilde} maximizes the differential entropy for a given Kullback-Leibler divergence to $P_Z$.
\item \label{it:lem4} For $\gamma\downarrow 0$,
\begin{IEEEeqnarray}{rCl}
{D}(P_{\tilde{Z}}\|P_Z) & = & \frac{\gamma^2}{2} \textnormal{Var} \left[ \ln(p_Z(Z))\right] + O(\gamma^3)\label{eq:TaylorD}\IEEEeqnarraynumspace \\
h(\tilde{Z}) - h(Z) & = & \gamma\,\textnormal{Var} \left[ \ln(p_Z(Z))\right] + O(\gamma^2). \label{eq:Taylorh}
\end{IEEEeqnarray}
\item \label{it:lem3} The function $\gamma\mapsto {D}(P_{\tilde{Z}}\|P_Z)$ is continuous on $[0,1-\zeta)$.
\end{enumerate}
\end{lemma}
\begin{IEEEproof}
We first prove \ref{it:lem1}) as follows:
\begin{IEEEeqnarray}{rCl}
\IEEEeqnarraymulticol{3}{l}{
D(P_{\tilde{Z}}\| P_Z)} \nonumber\\*
\label{computation_h_z_tilde}
\,\,\,\,\,&=&\int_{\mathbb{R}}p_{\tilde{Z}}(z)\ln\left(\frac{p_{\tilde{Z}}(z)}{p_{Z}(z)}\right) \dd z\nonumber\\
    &=&-h(\tilde{Z})-\int_{\mathbb{R}}p_{\tilde{Z}}(z)\ln(p_{Z}(z)) \dd z\nonumber\\
    &=&-h(\tilde{Z}) -\frac{1}{1-\gamma}\int_{\mathbb{R}}p_{\tilde{Z}}(z)\ln(\alpha p_{Z}(z)^{1-\gamma}) \dd z +\frac{\ln(\alpha)}{1-\gamma}\nonumber\\
    &=& -h(\tilde{Z}) + \frac{1}{1-\gamma} h(\tilde{Z}) + \frac{\ln(\alpha)}{1-\gamma}\nonumber\\
    &=& \frac{\gamma}{1-\gamma} \, h(\tilde{Z})+\frac{\ln(\alpha)}{1-\gamma} \label{eq:20}
\end{IEEEeqnarray}
which implies \eqref{entropy_z_tilde}.

We next show \ref{it:lem2}). For any random variable $Y$ satisfying \eqref{eq:Deps}, we have
\begin{IEEEeqnarray}{rCl}
\label{maximisation_entropy}
    0&\leq&{D}(P_{Y}\|P_{\tilde{Z}})\nonumber\\
    &=&-h(Y)-\int_{\mathbb{R}}p_{Y}(y)\ln\left(p_{\tilde{Z}}(y)\right)\textnormal{d}y\nonumber\\
    &=&-h(Y)-\int_{\mathbb{R}}p_{Y}(y)\ln\left(\alpha p_Z(y)^{1-\gamma}\right)\dd y\nonumber\\
    &=&-h(Y)-\ln(\alpha)-(1-\gamma)\int_{\mathbb{R}} p_Y(y)\ln\left(p_Z(y)\right)\dd y \nonumber\\
    &=&-h(Y)-\ln(\alpha) + (1-\gamma) {D}(P_Y\| P_Z)+ (1-\gamma) h(Y) \nonumber\\
    & \leq & -\gamma\, h(Y) - \ln(\alpha) + (1-\gamma) {D}(P_{\tilde{Z}} \| P_Z) \label{eq:21}\\
    &=&\gamma \bigl(h(\tilde{Z})-h(Y)\bigr), \label{eq:22}
\end{IEEEeqnarray}
where \eqref{eq:21} follows from \eqref{eq:Deps}; and \eqref{eq:22} from \eqref{eq:20}. Inequality \eqref{eq:22} implies \eqref{entropy_upper_bound}.

We next show \ref{it:lem4}). 
There exists $\theta\colon\mathbb{R}\to(0,\gamma)$ such that the Taylor expansion of $p_{Z}(z)^{-\gamma}$ with the Lagrange form of the remainder is
\begin{IEEEeqnarray}{rCl}
\label{limited_development_density_lemme}
    p_{Z}(z)^{-\gamma}&=&1-\gamma\ln(p_Z(z))+\frac{\gamma^2}{2}\left(\ln(p_Z(z))\right)^2\nonumber\\
    &&{}-\frac{\gamma^3}{6}\left(\ln(p_Z(z))\right)^3 p_Z(z)^{-\theta(z)}~~~~ z\in\mathbb{R}. \IEEEeqnarraynumspace
\end{IEEEeqnarray}
The normalization factor $\alpha$ is then
\begin{IEEEeqnarray}{rCl}
\label{limited_development_alpha_n}
    \alpha&=&\Bigg(\int_{\mathbb{R}}p_Z(z) \Bigg(1-\gamma\ln(p_Z(z))+\frac{\gamma^2}{2}\left(\ln(p_Z(z))\right)^2\nonumber\\
    &&\qquad\qquad\quad\quad{}-\frac{\gamma^3}{6}\left(\ln(p_Z(z))\right)^3 p_Z(z)^{-\theta(z)}\Bigg)\dd z\Bigg)^{-1}\nonumber \\
\label{dl_alpha_lemme}
 & = &\left(1+\gamma h(Z) + \frac{\gamma^2}{2}\mathbb{E}\left[\left(\ln(p_Z(Z))\right)^2\right] +O(\gamma^3)\right)^{-1}\IEEEeqnarraynumspace \\
 \label{eq:alpha22}
    &=&1-\gamma h(Z) - \frac{\gamma^2}{2}\mathbb{E}\left[\left(\ln(p_Z(Z))\right)^2\right] + \gamma^2 h(Z)^2 \nonumber\\
    &&{}+O(\gamma^3), 
\end{IEEEeqnarray}
where \eqref{dl_alpha_lemme} follows by Lemma~\ref{lemma_integrability} and \eqref{eq:alpha22} follows by the Taylor expansion of $x\mapsto (1+x)^{-1}$. Note that
\begin{IEEEeqnarray}{rCl}
\ln \alpha&=&-\ln\left(1+\gamma h(Z)+\frac{\gamma^2}{2}\mathbb{E}\left[\left(\ln(p_Z(Z))\right)^2\right]+O(\gamma^3)\right)\nonumber\\
&=&-\gamma h(Z)-\frac{\gamma^2}{2}\mathbb{E}\left[\left(\ln(p_Z(Z))\right)^2\right]+\frac{\gamma^2}{2}h(Z)^2+O(\gamma^3)\nonumber\\
&=&-\gamma h(Z)-\frac{\gamma^2}{2}\mathrm{Var}\left[\ln(p_Z(Z))\right]+O(\gamma^3).
\label{eq:logalpha}
\end{IEEEeqnarray}
We also have
\begin{IEEEeqnarray}{rCl}
\IEEEeqnarraymulticol{3}{l}{
\int_{\mathbb{R}}p_{Z}(z)^{1-\gamma}\ln(p_Z(z))\dd z
}\nonumber\\* ~~~~
\label{dl_int_pz_lemme}
    &=&\int_{\mathbb{R}}p_{Z}(z)\Bigg(1-\gamma\ln(p_Z(z))+\frac{\gamma^2}{2}\left(\ln(p_Z(z))\right)^2\nonumber\\
    &&\qquad\quad{}-\frac{\gamma^3}{6}\left(\ln(p_Z(z))\right)^3p_Z(z)^{-\theta(z)}\Bigg)\ln(p_Z(z))\dd z\nonumber\\
    &=&-h(Z)-\gamma\mathbb{E}\left[\left(\ln(p_Z(Z))\right)^2\right]+\frac{\gamma^2}{2}\mathbb{E}\left[\left(\ln(p_Z(Z))\right)^3\right]\nonumber\\* && {}+O(\gamma^3).
\end{IEEEeqnarray}
(The expectations above are finite by Lemma~\ref{lemma_integrability}.) We can now compute the Taylor expansion of ${D}(P_{\tilde{Z}}\|P_{Z})$ in $\gamma$:
\begin{IEEEeqnarray}{rCl}
    \IEEEeqnarraymulticol{3}{l}{{D}(P_{\tilde{Z}}\|P_{Z})}\nonumber\\*
    &=&\int_{\mathbb{R}}p_{\tilde{Z}}(z)\ln\left(\alpha \frac{p_Z(z)^{1-\gamma}}{p_Z(z)}\right)\dd z\nonumber\\
    &=&\ln(\alpha)-\gamma\alpha\int_{\mathbb{R}}p_{Z}(z)^{1-\gamma}\ln(p_Z(z))\dd z\nonumber\\
    &=&-\gamma h(Z)-\frac{\gamma^2}{2}\mathbb{E}\left[\left(\ln(p_Z(Z))\right)^2\right]+\frac{\gamma^2}{2}h(Z)^2+O(\gamma^3)\nonumber\\*
    &&{}-\gamma\Bigg(1-\gamma h(Z) - \frac{\gamma^2}{2}\mathbb{E}\left[\left(\ln(p_Z(Z))\right)^2\right] + {\gamma^2} h(Z)^2 \nonumber\\*
    && \qquad\quad{}+O(\gamma^3)\Bigg)
    \cdot \Bigg({-}h(Z)-\gamma\mathbb{E}\left[\left(\ln(p_Z(Z))\right)^2\right] \nonumber\\
    &&~~~~~~~~~~~~~~~~~~~~ {}+\frac{\gamma^2}{2}\mathbb{E}\left[\left(\ln(p_Z(Z))\right)^3\right]+O(\gamma^3)\Bigg)   \label{bound_ln_2_3_4_lemme_use} \IEEEeqnarraynumspace \\
    &=&-\gamma h(Z)-\frac{\gamma^2}{2}\mathbb{E}\left[\left(\ln(p_Z(Z))\right)^2\right]+\frac{\gamma^2}{2}h(Z)^2+\gamma h(Z)\nonumber\\*
    &&{}+\gamma^2\mathbb{E}\left[\left(\ln(p_Z(Z))\right)^2\right] -\gamma^2h(Z)^2+O(\gamma^3)\nonumber\\
    &=&\frac{\gamma^2}{2}\mathbb{E}\left[\left(\ln(p_Z(Z))\right)^2\right]-\frac{\gamma^2}{2}h(Z)^2+O(\gamma^3)\nonumber\\
    \label{equation_divergence_generalized_case_lemme}
    &=&\frac{\gamma^2}{2}\mathrm{Var}\left[\ln(p_Z(Z))\right]+O(\gamma^3), 
\end{IEEEeqnarray}
where \eqref{bound_ln_2_3_4_lemme_use} follows by \eqref{dl_alpha_lemme}, \eqref{eq:alpha22}, \eqref{eq:logalpha} and \eqref{dl_int_pz_lemme}.
Similarly we compute the Taylor expansion of $h(\tilde{Z})-h(Z)$: using \eqref{entropy_z_tilde} we find
\begin{IEEEeqnarray}{rCl}
    h(\tilde{Z})-h(Z)&=& -\frac{\ln (\alpha)}{\gamma} + \frac{1-\gamma}{\gamma} \,{D}(P_{\tilde{Z}}\| P_Z)-h(Z)\nonumber\\
    \label{second_use_lagrange_form}
    &=& \frac{\gamma h(Z)+\frac{\gamma^2}{2}\mathrm{Var}\left[\ln(p_Z(Z))\right]+O(\gamma^3)}{\gamma}\nonumber\\
    &&{}+ \frac{1-\gamma}{\gamma} \left(\frac{\gamma^2}{2}\mathrm{Var}\left[\ln(p_Z(Z))\right]+O(\gamma^3)\right) \nonumber\\
    &&{}-h(Z) \\
    \label{use_of_lemma}
    &=& \gamma\mathrm{Var}\left[\ln(p_Z(Z))\right]+O(\gamma^2),
\end{IEEEeqnarray}
where \eqref{second_use_lagrange_form} follows by \eqref{eq:logalpha} and \eqref{equation_divergence_generalized_case_lemme}.

Finally, we show \ref{it:lem3}). Continuity at $\gamma=0$ follows by \eqref{equation_divergence_generalized_case_lemme}, because the latter implies
\begin{equation}
    \lim\limits_{\gamma\downarrow 0}{D}(P_{\tilde{Z}}\|P_Z) =0.
\end{equation}
We next write
\begin{IEEEeqnarray}{rCl}
    {D}(P_{\tilde{Z}}\|P_Z)&=&\int_{\mathbb{R}}\alpha p_Z(z)^{1-\gamma}\ln\left(\frac{\alpha p_Z(z)^{1-\gamma}}{p_Z(z)}\right)\dd z\nonumber\\
    &=&-\gamma\alpha\int_{\mathbb{R}} p_Z(z)^{1-\gamma}\ln(p_Z(z))\dd z+\ln(\alpha). \IEEEeqnarraynumspace
\end{IEEEeqnarray}
To prove the desired continuity, it suffices to show that both $\alpha$ and 
\begin{equation}\label{eq:intint}
    \int_{\mathbb{R}}p_Z(z)^{1-\gamma}\ln(p_Z(z))\dd z
\end{equation} 
are continuous in $\gamma\in(0,1-\zeta)$. The statement for $\alpha$ follows by \eqref{eq:alpha22}. 
For $k=0,1$ the function
\begin{equation}
    \gamma\mapsto p_Z(z)^{1-\gamma}\ln(p_Z(z))^k
\end{equation}
is continuous for every $z\in\mathbb{R}$. 
For all $\gamma \in (0, 1-\zeta)$, the function
\begin{equation}
    z\mapsto p_Z(z)^{1-\gamma}\ln(p_Z(z))^k
\end{equation}
is integrable by Lemma~\ref{lemma_integrability}. Furthermore, 
\begin{IEEEeqnarray}{rCl}
\IEEEeqnarraymulticol{3}{l}{
\left|p_Z(z)^{1-\gamma}\left(\ln(p_Z(z))\right)^k\right|
}\nonumber\\* ~~~~~
    &\leq& p_Z(z)\left|\ln(p_Z(z))\right|^k+p_Z(z)^{\zeta}\left|\ln(p_Z(z))\right|^k, \IEEEeqnarraynumspace
\end{IEEEeqnarray}
where the right-hand side is again integrable by Lemma~\ref{lemma_integrability}. 
By the lemma of continuity under integrals 
\cite[11.4]{schilling2017measures},
we conclude that both $\alpha$ and \eqref{eq:intint} are continuous in $\gamma\in(0,1-\zeta)$, hence so is ${D}\left(P_{\tilde{Z}}\|P_{Z}\right)$.
\end{IEEEproof}

\vspace{2mm}

\begin{IEEEproof}[Proof of Theorem \ref{theorem_converse}]
Take any 
code $\mathcal{C}$ of length $n$.
Let $\bar{X}$ denote a random variable such that $P_{\bar{X}}$ is the average input distribution over the secret key,
a uniformly drawn message, and the $n$ channel uses, and let $\bar{Y}$ denote the channel output random variable when the input is $\bar{X}$, so $P_{\bar{Y}}$ is the average output distribution (in the same sense as $P_{\bar{X}}$):
\begin{IEEEeqnarray}{rCl}
    P_{\bar{X}}(\cdot) & = \frac{1}{n}\sum_{i=1}^n P_{X_i}(\cdot),\\
    P_{\bar{Y}}(\cdot) & = \frac{1}{n}\sum_{i=1}^n P_{Y_i}(\cdot).
\end{IEEEeqnarray}

Starting with the condition \eqref{covert_communication_hypothesis}, similarly to \cite{fundamental_covertness} we have:
\begin{IEEEeqnarray}{rCl}
\label{block_chain_divergence}
    \Delta&\geq&{D}(P_{Y^n}\|P_{Z^n})\nonumber\\
    &=&-h(Y^n)-\mathbb{E}[\ln \left(p_{Z^n}(Y^n)\right)]\nonumber\\
    &=&\sum_{i=1}^n\left(-h(Y_i|Y^{i-1})-\mathbb{E}[\ln(p_{Z}(Y_i))]\right)\nonumber\\
    &\geq&\sum_{i=1}^n\left(-h(Y_i)-\mathbb{E}[\ln(p_{Z}(Y_i))]\right)\nonumber\\
    &=&\sum_{i=1}^n{D}(P_{Y_i}\|P_{Z})\nonumber \\
    \label{eq:DYbarZ}
    &\geq& n\,{D}(P_{\bar{Y}}\|P_{Z}),
\end{IEEEeqnarray}
where the last step follows because the Kullback-Leibler divergence is convex.

We next derive a bound on $A_n(\Delta,\epsilon_n)$ in terms of $\bar{X}$ and $\bar{Y}$. For each realization 
$\mathsf{K}=k$ of the secret key,
we denote by 
$\epsilon_{n,k}$ 
the error probability of the code $\mathcal{C}$ with $\mathsf{K}=k$.
Let $\epsilon_n$ be the 
error probability averaged over the key.
For each 
$k \in \mathcal{K}$,
we have by Fano's inequality:
\begin{equation}   \ln\left|\mathcal{M}\right|(1-\epsilon_{n,k}
)- 1  \leq I(X^n;Y^n|
\mathsf{K}=k),
\end{equation}
where the joint distribution on $(X^n,Y^n)$ is computed according to a uniformly drawn message.
By averaging over the key,
we obtain
\begin{IEEEeqnarray}{rCl}
   \ln\left|\mathcal{M}\right|(1-\epsilon_n)- 1
    &\leq& I(X^n;Y^n|\mathsf{K})\nonumber\\ 
    &\leq& I(X^n,\mathsf{K}; Y^n)\nonumber\\
    \label{markov_chain_inequality}
    &=&I(X^n;Y^n)\\ 
    &=&\sum_{i=1}^n I(X^n ; Y_i |Y^{i-1} )\nonumber\\
    &=&\sum_{i=1}^n \big(h(Y_i |Y^{i-1} )-h(Y_i |X^n, Y^{i-1} )\big)\nonumber \\
  &=&\sum_{i=1}^n \left(h(Y_i|Y^{i-1}) - h(Y_i |X_i)\right)\nonumber\\
  \label{mutual_information_channel_one_shot}
  &\leq&\sum_{i=1}^n I(X_i;Y_i)\nonumber \\
  \label{mutual_information_upper_bound_first_tmp}
  &\leq& nI(\bar{X};\bar{Y})\\
  & = & n(h(\bar{Y}) - h(Z)),\label{eq:39}
\end{IEEEeqnarray}
where \eqref{markov_chain_inequality} holds by the Markov chain:
\begin{equation}
\label{markov_chain}
    \mathsf{K}\rightarrow X^n \rightarrow Y^n; 
\end{equation}
and \eqref{mutual_information_upper_bound_first_tmp} because mutual information is concave in the input distribution.
By the definition of $A_n(\Delta,\epsilon)$, \eqref{eq:39} implies 
\begin{IEEEeqnarray}{rCl}
\label{mutual_information_upper_bound_first}
    A_n(\Delta,\epsilon_n)(1-\epsilon_n)- 1 &\leq& n(h(\bar{Y})-h(Z)).
\end{IEEEeqnarray}

We shall complete the proof using \eqref{eq:DYbarZ}, \eqref{mutual_information_upper_bound_first}, and Lemma~\ref{lemma_inequalitiy_entropy_generalized}. We first treat the special case where $Z$ has the uniform distribution over a subset of $\mathbb{R}$; let us denote this subset by $\mathcal{S}$, so
\begin{equation}\label{eq:uniform}
p_Z(z) = \begin{cases} \frac{1}{\lambda\left(\mathcal{S}\right)}, & z\in\mathcal{S}\\ 0,& \textnormal{otherwise,} \end{cases}
\end{equation}
where $\lambda$ denotes the Lebesgue measure.
In this case, one can show that covert communication is not possible at all, so $L=0$; we provide a proof in Appendix~\ref{appendix_uniform_case}. It then follows that \eqref{eq:upper} trivially holds in this case.

In the rest of this proof we shall assume that $p_Z$ does not have the form \eqref{eq:uniform}. 
Then, given any $\gamma>0$, $P_{\tilde{Z}}$ defined by \eqref{def_z_tilde} differs from $P_Z$, therefore ${D}(P_{\tilde{Z}}\|P_{Z})>0$. 
Recall that the function $\gamma \mapsto {D}(P_{\tilde{Z}}\|P_{Z})$ is continuous on $[0,1-\zeta)$; see Lemma~\ref{lemma_inequalitiy_entropy_generalized}, part \ref{it:lem3}). Therefore, there exists a sequence $\left\{\gamma_n\right\}$ such that
\begin{equation}
\lim_{n\to\infty} \gamma_n=0,
\end{equation}
and, for large enough $n$, the PDFs defined as
\begin{equation}
\label{def_z_tilde_n}
    p_{\tilde{Z}_n}(\tilde{z}) =\alpha_n \cdot p_Z(\tilde{z})^{1-\gamma_n}, \quad \tilde{z} \in \mathbb{R},
\end{equation}
with
\begin{equation}
    \alpha_n = \left(\int_{\mathbb{R}}p_Z(z)^{1-\gamma_n} \dd z\right)^{-1},
\end{equation}
satisfy
\begin{equation}
\label{def_lambda_by_continuity}
{D}(P_{\tilde{Z}_n}\|P_Z)=\frac{\Delta}{n}.
\end{equation}
Using \eqref{eq:TaylorD} we then have
\begin{equation}
    \frac{\Delta}{n} = \frac{\gamma_n^2}{2}\mathrm{Var}\left[\ln(p_Z(Z))\right]+O(\gamma_n^3),
\end{equation}
which implies
\begin{equation}
    \label{lambda_upper_bound}
    \gamma_n = \sqrt{\frac{2}{\mathrm{Var}\left[\ln(p_Z(Z))\right]}}\sqrt{\frac{\Delta}{n}}+o\left(\frac{1}{\sqrt{n}}\right).
\end{equation}
We now continue from \eqref{mutual_information_upper_bound_first} as follows:
\begin{IEEEeqnarray}{rCl}
\IEEEeqnarraymulticol{3}{l}{
\frac{ A_n(\Delta,\epsilon_n) (1-\epsilon_n)-1}{n}
}\nonumber\\* ~~~~~~~~~~
 & \le & h(\bar{Y}) - h(Z) \nonumber\\
& \le & h(\tilde{Z}_n) - h(Z) \label{eq:ZnZ}\\
& = & \gamma_n\mathrm{Var}\left[\ln(p_Z(Z))\right]+O(\gamma_n^2) \label{eq:Varn}\\
& = & \sqrt{2}\sqrt{\mathrm{Var}\left[\ln(p_Z(Z))\right]}\sqrt{\frac{\Delta}{n}} + o\left(\frac{1}{\sqrt{n}}\right),\label{eq:upperfinal} \IEEEeqnarraynumspace
\end{IEEEeqnarray}
where \eqref{eq:ZnZ} follows by \eqref{entropy_upper_bound}, \eqref{eq:DYbarZ}, and \eqref{def_lambda_by_continuity}; \eqref{eq:Varn} by \eqref{eq:Taylorh}; and \eqref{eq:upperfinal} by \eqref{lambda_upper_bound}. Recalling the definition \eqref{L_definition} and taking $n\rightarrow\infty$ and $\epsilon_n\rightarrow 0$ in \eqref{eq:upperfinal} complete the proof.
\end{IEEEproof}

\begin{remark}
Theorem \ref{theorem_converse} is reminiscent of \cite[Theorem 3]{fundamental_covertness}, which provides an upper bound on $L$ for DMCs in terms of the output distribution induced by the ``off'' input symbol, denoted $Q_0$, and the capacity-achieving output distribution, denoted $Q^*$. If $Q^*$ is the uniform distribution, then \cite[Theorem 3]{fundamental_covertness} has a similar form to Theorem~\ref{theorem_converse} with the PDF $p_Z$ replaced by the discrete distribution~$Q_0$. 
\end{remark}

\begin{remark}
The right-hand side of \eqref{eq:upper} does not change when the noise $Z$ is scaled by a constant factor. This is true in general: for any channel \eqref{eq:channel}, scaling the additive noise by a constant factor $c$ does not affect $L$, because this effect can be canceled out by multiplying the input $X$ by $c$ at the transmitter. Then, for both the receiver and the eavesdropper, there is no loss in optimality in scaling the output by $1/c$ to recover the same $Y$ as in the original channel.
\end{remark}

\section{Tightness of the Upper Bound}
\label{section:lower_bound_L}

Under some additional assumptions on the noise PDF $p_Z$, one can show that the upper bound of Theorem~\ref{theorem_converse} is tight. 

\begin{assumption}\label{assum}
Assume that $p_Z$ satisfies the following:
\begin{enumerate}
    \item \label{item:ass1}$p_Z$ is bounded, i.e., there exists some $b>0$ such that
  \begin{equation}
  \label{eq:bounded}
      p_Z(z) \leq b,\quad  z\in\mathbb{R};
  \end{equation}
\item \label{item:ass2} $z\mapsto p_Z(z)\ln\left(p_Z(z)\right)$ is uniformly continuous on its support $\textnormal{supp}(p_Z)$, i.e., for all $\epsilon>0$, there exists $\delta>0$ such that, for any $z_1,z_2\in\textnormal{supp}(p_Z)$, $\left|z_1-z_2\right|\leq\delta$, we have 
\begin{IEEEeqnarray}{rCl}
\label{assumption_2}
\bigl|p_{Z}(z_1)\ln(p_Z(z_1))-p_{Z}(z_2)\ln(p_Z(z_2)) \bigr|&\leq&\epsilon; \IEEEeqnarraynumspace
\end{IEEEeqnarray}
\item \label{item:ass3} there exists some $\xi\in(0,1)$ such that, for all $\gamma\in[0,\xi)$, there exists a random variable $X$ independent of $Z\sim p_Z$ such that
   the PDF of $X+Z$ is $p_{\tilde{Z}}$ given by \eqref{def_z_tilde}.
   \end{enumerate}
\end{assumption}

\begin{theorem}
\label{theorem_information_spectrum}
For the memoryless additive-noise channel \eqref{eq:channel}, if $p_Z$ satisfies \eqref{eq:integrable0}--\eqref{eq:integrable2} as well as Assumption~\ref{assum}, then
\begin{equation} \label{eq:Lequal}
    L = \sqrt{2} \sqrt{ \textnormal{Var} \left[\ln(p_Z(Z))\right]}.
\end{equation}
\end{theorem}

Before proving Theorem \ref{theorem_information_spectrum}, we
recall L\'evy's convergence theorem concerning \emph{weak convergence} (also called \emph{convergence in distribution}) of real-valued random variables,\footnote{Weak convergence can also be defined for general probability measures, but that is not needed in the present work.} and prove a lemma.

\begin{definition}[Characteristic function {\cite[16.1]{williams_probability_2018}}]
Let $X$ be a real-valued random variable. The characteristic function of $X$, $\varphi_X\colon \mathbb{R}\to\mathbb{C}$, $t\mapsto \varphi_X(t)$, is given by
\begin{equation}
    \varphi_X(t) = \mathbb{E} \left[ e^{i t X}\right],\quad t\in\mathbb{R}.
\end{equation}
\end{definition}

\begin{definition}[Weak convergence {\cite[17.1]{williams_probability_2018}}]
Let $\{X_n\}$ be a sequence of random variables and $X$ be another random variable. The probability distributions $\left\{P_{X_n}\right\}$ converge weakly to $P_X$ as $n\rightarrow \infty$ if, for every bounded continuous function $f$ on $\mathbb{R}$,
\begin{equation}
\lim_{n\to\infty} \mathbb{E} \left[ f(X_n) \right] = \mathbb{E} \left[ f(X)\right].
\end{equation}
\end{definition}

Since both the real and the imaginary parts of the function $x\mapsto e^{itx}$ are bounded and continuous for every $t\in\mathbb{R}$, if $\{P_{X_n}\}$ converge weakly to $P_X$ as $n\to\infty$, then the characteristic functions of $\{X_n\}$ must converge pointwise to the characteristic function of $X$. The reverse is also true:

\begin{theorem}[L\'evy's convergence theorem {\cite[18.1]{williams_probability_2018}}]
\label{levy_theorem}
Consider a sequence of random variables $\{{X_n}\}$ with respective characteristic functions $\{\varphi_{X_n}\}$. If, as $n\to\infty$, $\varphi_{X_n}$ converges pointwise to some function $\varphi$ that is continuous at $0$, then $\varphi$ is the characteristic function of some random variable $X$, and $P_{X_n}$ converges weakly to $P_X$.
\end{theorem}

Let $P_0$ denote the degenerate (i.e., deterministic) distribution
\begin{equation}
\label{dirac}
    P_0 (x) =\begin{cases} 0,&x<0\\ 1, & x\ge 0,\end{cases}
\end{equation}
and let $\varphi_0$ denote the corresponding characteristic function, so
\begin{equation}
\varphi_0(t)=1,\quad t\in\mathbb{R}.
\end{equation}

\begin{lemma}
\label{weak_convergence_p_xn}
Consider any $p_Z$ satisfying \eqref{eq:integrable0}--\eqref{eq:integrable2} and Assumption~\ref{assum}. For every $n$, let $\gamma_n \in (0, \xi)$, and let $X_n$ be independent of $Z$ and be such that
\begin{equation}\label{eq:ZnXnZ}
    \tilde{Z}_n=X_n+Z
\end{equation}
has density
\begin{equation}
\label{def_z_tilde_lemma_weak_convergence}
    p_{\tilde{Z}_n}(\tilde{z})=\alpha_n p_Z(\tilde{z})^{1-\gamma_n}, \quad \tilde{z} \in \mathbb{R},
\end{equation}
where
\begin{equation}
    \alpha_n = \left(\int_{\mathbb{R}}p_{Z}(z)^{1-\gamma_n} \dd z\right)^{-1}.
\end{equation}
If $\gamma_n\to 0$ as $n\to\infty$, then,  as $n\to\infty$, $P_{\tilde{Z}_n}$ converges weakly to $P_Z$, and $P_{X_n}$ converges weakly to $P_0$.
\end{lemma}
\begin{IEEEproof}
We first show that $P_{\tilde{Z}_n}$ converges weakly to $P_Z$. To this end, take any bounded continuous function $f$ on $\mathbb{R}$, and we have
\begin{IEEEeqnarray}{rCl} 
\IEEEeqnarraymulticol{3}{l}{
\left| \mathbb{E}\left[f(\tilde{Z}_n)\right] - \mathbb{E} [f(Z)] \right|
}\nonumber\\* ~~~~~~~~~~
 & = & \left|\int_{\mathbb{R}}f(z)p_{\tilde{Z}_n}(z)\dd z-\int_{\mathbb{R}}f(z)p_Z(z)\dd z\right|\nonumber\\
&\leq&\|f\|_{\infty}\int_{\mathbb{R}}\left|p_{\tilde{Z}_n}(z)-p_Z(z)\right|\dd z\nonumber\\    &=&\|f\|_{\infty}\left\|P_{\tilde{Z}_n}-P_{Z}\right\|_{1}\nonumber\\
    \label{use_pinsker_inequality}   &\leq&\|f\|_{\infty}\sqrt{2{D}\left(P_{\tilde{Z}_n}\|P_{Z}\right)},
\end{IEEEeqnarray}
where \eqref{use_pinsker_inequality} follows by Pinsker's inequality. 
The right-hand side of \eqref{use_pinsker_inequality} tends to $0$ as $n\to\infty$, because $\gamma_n\to 0$ and by Lemma~\ref{lemma_inequalitiy_entropy_generalized}. It follows that $\mathbb{E}[f(\tilde{Z}_n)]$ tends to $\mathbb{E}[f(Z)]$ as $n\to\infty$, therefore, by definition, $P_{\tilde{Z}_n}$ converges weakly to $P_Z$. 

The above implies
\begin{equation}\label{eq:limvarphi}
    \lim\limits_{n\rightarrow \infty}\varphi_{\tilde{Z}_n}(t)=\varphi_{Z}(t),\quad t\in\mathbb{R}.
\end{equation}
For any $n$, since $X_n$ is independent of $Z$, we have \cite[16.4]{williams_probability_2018}
\begin{equation}\label{eq:varphiprod}
    \varphi_{\tilde{Z}_n}(t)=\varphi_{Z}(t) \varphi_{X_n}(t), \quad t\in\mathbb{R},
\end{equation}
where $\varphi_{X_n}$ is the characteristic function of $X_n$. 
By properties of characteristic functions \cite[16.2]{williams_probability_2018}, $\varphi_{Z}$ is continuous and $\varphi_{Z}(0)=1$, hence there exists an interval around $0$ on which $\varphi_{Z}(t)\neq 0$. By \eqref{eq:limvarphi} and \eqref{eq:varphiprod}, for any $t$ in this interval, 
\begin{equation}\label{eq:limisone}
    \lim_{n\to \infty}\varphi_{X_n}(t)=1.
\end{equation}
By, e.g., \cite[Theorem~9.6.4]{fourier_1972}, we know that \eqref{eq:limisone} must hold for all $t\in\mathbb{R}$. This and Theorem~\ref{levy_theorem} imply that $P_{X_n}$ converges weakly to $P_0$ as $n\to\infty$.
\end{IEEEproof}

\begin{IEEEproof}[Proof of Theorem \ref{theorem_information_spectrum}]
\label{proof_theorem_2}
The converse part follows immediately from Theorem~\ref{theorem_converse}. We shall prove the direct part using techniques similar to \cite[Theorem 3]{bouette2023covert}, \cite[Section V-B]{fundamental_covertness}, together with Lemma~\ref{weak_convergence_p_xn}.

Fix $\chi\in(1,\frac{3}{2})$. For sufficiently large $n$, let $\tilde{Z}_n$ have the PDF \eqref{def_z_tilde_lemma_weak_convergence}, with the choice
\begin{equation}\label{eq:gamman}
\gamma_n=\sqrt{\frac{2}{\mathrm{Var}\left[\ln(p_Z(Z))\right]}\left(\frac{\Delta}{n}-\frac{1}{n^{\chi}}\right)}.
\end{equation}
(It will become clear later on that this choice of $\gamma_n$ is to satisfy the covertness condition.)
Further, let $X$ be independent of $Z$ and have the distribution of $X_n$ satisfying \eqref{eq:ZnXnZ}; the existence of such $X$ (for sufficiently large $n$) is guaranteed by Assumption~\ref{assum}. 

We generate a random codebook $\mathsf{C}$ by picking every codeword i.i.d. according to $P_{X}$ and independently of the other codewords. 
Note that
\begin{IEEEeqnarray}{rCl}
\IEEEeqnarraymulticol{3}{l}{\mathbb{E}_{\mathsf{C}}[{D}(P_{Y^n|\mathsf{C}}\|P_{Z^n})]}\nonumber\\*
\qquad&=&\mathbb{E}_{\mathsf{C}}\left[{D}\left(P_{Y^n|\mathsf{C}}\middle\|P_{\tilde{Z}_n}^{\times n}\right)\right]+{D}\left(P_{\tilde{Z}_n}^{\times n}\middle\|P_{Z^n}\right). \IEEEeqnarraynumspace
\label{eq:covertness2}
\end{IEEEeqnarray}
For the moment we assume the key is arbitrarily long and therefore allows the output $P_{Y^n|\mathsf{C}}$ of the random codebook averaged over the key to be ``sufficiently close'' to i.i.d. with high probability. This means that the first term in \eqref{eq:covertness2} is close to zero. We admit that this assumption is unrealistic, but 
we shall show in Section \ref{section:key_length} that a finite key would suffice. 
To show that $\mathsf{C}$ satisfies the covertness condition \eqref{covert_communication_hypothesis} with high probability, it now suffices to show that the second term on the right-hand side of \eqref{eq:covertness2} is less than $\Delta$ for sufficiently large $n$. 
This can be shown as follows:
\begin{IEEEeqnarray}{rCl}
    \label{iid_generation}
    {D}\left(P_{\tilde{Z}_n}^{\times n}\middle\|P_{Z^n}\right)&=&n\,{D}\left(P_{\tilde{Z}_n}\middle\|P_{Z}\right)\\
    \label{check_satisfaction_covertness_constraint}
    &=&n \frac{\gamma_n^2}{2}\mathrm{Var}\left[\ln(p_Z(Z))\right]+O(n \gamma_n^3)\IEEEeqnarraynumspace \\
    \label{check_satisfaction_covertness_constraint_2}
    &=&\Delta-n^{1-\chi}+O\left(\frac{1}{\sqrt{n}}\right),
\end{IEEEeqnarray}
where 
\eqref{check_satisfaction_covertness_constraint} follows by \eqref{eq:TaylorD}; and \eqref{check_satisfaction_covertness_constraint_2} by \eqref{eq:gamman}. From \eqref{check_satisfaction_covertness_constraint_2} it is clear that 
${D}(P_{\tilde{Z}_n}^{\times n}\|P_{Z^n}) < \Delta$ 
for sufficiently large $n$.

We next look at the mutual information corresponding to this code construction. To simplify notation, we henceforth use $Y$ to denote the single-letter channel output, i.e., $Y=X+Z$; recall that $Y \sim \tilde{Z}_n$ and that the distributions of $X$ and $Y$ both depend on $n$.  
Since the inputs are i.i.d., we have
\begin{IEEEeqnarray}{rCl}
\lim_{n\to\infty} \frac{1}{\sqrt{n}} I(X^n;Y^n) & = & \lim_{n\to\infty} \sqrt{n}\, I(X;Y) \nonumber\\
& = & \lim_{n\to\infty} \sqrt{n}\, \bigl( h(Y) - h(Z)\bigr) \nonumber  \\
& = & \lim_{n\to\infty} \sqrt{n}\, \left( h(\tilde{Z}_n) - h(Z)\right)\nonumber \\
& = & \sqrt{2}\sqrt{\textnormal{Var}[\ln(p_Z(Z))]} \cdot \sqrt{\Delta}, \label{eq:equalVar} \IEEEeqnarraynumspace
\end{IEEEeqnarray}
where the last step follows by \eqref{eq:Taylorh} and \eqref{eq:gamman}.

It now remains to show that this limit of mutual information is operationally achievable. Namely, we wish to show
\begin{equation}
\label{lower_boud_L}
    \lim\limits_{\epsilon\downarrow  0}\liminf\limits_{n \rightarrow \infty} \frac{A_n(\Delta,\epsilon)}{\sqrt{n}} \geq \lim\limits_{n \rightarrow \infty} \frac{1}{\sqrt{n}} I(X^n;Y^n).
\end{equation}
From 
\cite[Theorem 18.5]{info_theory_polyanskiy}
we know that, for any $\tau>0$ 
the average error probability  $\epsilon$ of the above random code is bounded as
\begin{equation}
\label{feinstein_lemma}
    \epsilon\leq\mathbb{P}\left\{i_{X^n,Y^n}(X^n,Y^n) \leq \ln|\mathcal{M}_n| +\tau\right\}+e^{-\tau},
\end{equation}
where $\mathcal{M}_n$ denotes the message set to be transmitted when the total number of channel uses is $n$; and where
\begin{equation}
\label{definition_information_density}
    i_{X^n,Y^n}(x^n,y^n)=\ln\left(\frac{p_{Y^n|X^n}(y^n|x^n)}{p_{Y^n}(y^n)}\right)
\end{equation}
is the \emph{information density}.
Choosing $\tau=n^{\frac{1}{4}}$, we rewrite \eqref{feinstein_lemma} as
\begin{IEEEeqnarray}{c}
\label{feinstein_lemma_step}
    \epsilon \leq\mathbb{P}\left\{\frac{i_{X^n,Y^n}(X^n,Y^n)}{\sqrt{n}} \leq \frac{\ln|\mathcal{M}_n|}{\sqrt{n}} +n^{-\frac{1}{4}}\right\} + e^{-n^{\frac{1}{4}}}. \nonumber\\*
\end{IEEEeqnarray}
Letting $n\to\infty$ in \eqref{feinstein_lemma_step}, we know that, with high probability, the average error probability of the random code $\mathsf{C}$ will tend to zero provided
\begin{IEEEeqnarray}{rCl}
\label{fundamental_formula_capacity}
\lim_{n \rightarrow \infty} \frac{\ln|\mathcal{M}_n|}{\sqrt{n}}
     &<& \mathbb{P}\text{-}\liminf\limits_{n \rightarrow \infty} \frac{1}{\sqrt{n}} i_{X^n,Y^n}(X^n,Y^n), \IEEEeqnarraynumspace
\end{IEEEeqnarray}
where $\mathbb{P}\text{-}\liminf$ denotes the \textit{limit inferior in probability}; see \cite{general_capacity}.
We shall show that the term inside the $\mathbb{P}\text{-}\liminf$ converges in probability to its expectation 
\begin{equation}
\label{expection_information_density}
\mathbb{E}\left[\frac{1}{\sqrt{n}} i_{X^n,Y^n}(X^n,Y^n)\right] = \frac{1}{\sqrt{n}} I(X^n;Y^n),
\end{equation}
which, combined with \eqref{eq:equalVar}, will yield the desired achievability result. 
By Chebyshev's inequality, for any $a >0$,
\begin{IEEEeqnarray}{rCl}
\IEEEeqnarraymulticol{3}{l}{
\mathbb{P}\left\{\left|\frac{1}{\sqrt{n}} i_{X^n,Y^n}(X^n,Y^n)- \mathbb{E}\left[ \frac{1}{\sqrt{n}} i_{X^n,Y^n}(X^n,Y^n)\right]\right| \geq a\right\}
}\nonumber\\* ~~~~~~~~~~~~~~~~~~~~~~~
\label{chebyshev}
&\leq& \frac{\mathrm{Var}\left(\dfrac{1}{\sqrt{n}} i_{X^n,Y^n}(X^n,Y^n)\right)}{a^2}.
\end{IEEEeqnarray}
Hence, to prove the desired convergence in probability and thereby the desired achievability result, it now only remains to show that the variance on the right-hand side of \eqref{chebyshev} converges to $0$ as $n\to\infty$. 

Recall that $Y$ has the same PDF as $\tilde{Z}_n$, which has the following Taylor expansion
\begin{IEEEeqnarray}{rCl}
\label{other_dl}
p_Y(y) & = & p_{\tilde{Z}_n}(y) \nonumber\\
&= & \alpha_n p_{Z}(y)\left(1-\gamma_n\ln(p_Z(y))p_Z(y)^{-\theta_n(y)}\right), \IEEEeqnarraynumspace
\end{IEEEeqnarray}
where $\theta_n(y)\in(0,\gamma_n)$ for all $y\in\mathbb{R}$. 
We now bound the variance in \eqref{chebyshev} as follows:
\begin{IEEEeqnarray}{rCl}
\IEEEeqnarraymulticol{3}{l}{
\mathrm{Var}\left(\frac{1}{\sqrt{n}} i_{X^n,Y^n}(X^n,Y^n)\right)
}\nonumber\\* 
&=&\mathrm{Var}\left(\frac{1}{\sqrt{n}} \ln\left(\frac{p_{Y^n|X^n}(Y^n|X^n)}{p_{Y^n}(Y^n)}\right)\right)\nonumber\\
&=&\mathrm{Var}\left(\ln\left(\frac{p_{Y|X}(Y|X)}{p_{Y}(Y)}\right)\right)\label{eq:IID}\\
&=&\mathrm{Var}\left(\ln\left(\frac{p_Z(Z)}{p_{Y}(Y)}\right)\right)\nonumber\\
\label{use_of_dl}
&=&\mathrm{Var}\left(\ln\left(\frac{p_Z(Z)}{\alpha_n p_{Z}(Y)(1-\gamma_n\ln(p_Z(Y))p_Z(Y)^{-\theta_n(Y)})}\right)\right)\nonumber\\
&&~\\
&=&\mathrm{Var}\Bigg(\ln\left(\frac{p_Z(Z)}{p_{Z}(Y)}\right)\nonumber\\*
&&\quad\qquad {}-\ln\left(1-\gamma_n\ln(p_Z(Y))p_Z(Y)^{-\theta_n(Y)}\right)-\ln(\alpha_n)\Bigg)\nonumber\\
&\leq&\mathbb{E}\Bigg[\Bigg(\ln\left(\frac{p_Z(Z)}{p_{Z}(Y)}\right)\nonumber\\*
&&\qquad\quad{}-\ln\left(1-\gamma_n\ln(p_Z(Y))p_Z(Y)^{-\theta_n(Y)}\right) \Bigg)^2\Bigg]\nonumber\\
&\le&2\, \mathbb{E}\left[\left(\ln\left(\frac{p_Z(Z)}{p_{Z}(Y)}\right)\right)^2\right]\nonumber\\
&&{}+2\, \mathbb{E}\left[\left(\ln\left(1-\gamma_n\ln\left(p_Z(Y)\right)p_Z(Y)^{-\theta_n(Y)}\right)\right)^2\right],\IEEEeqnarraynumspace
\label{main_computation_variance}
\end{IEEEeqnarray}
where \eqref{eq:IID} follows because $(X^n,Y^n)$ are i.i.d.; \eqref{use_of_dl} by \eqref{other_dl}; and \eqref{main_computation_variance} by adding the nonnegative term
\begin{IEEEeqnarray*}{l}
\mathbb{E}\bigg[\bigg(\hspace{-0.2em}\ln\left(\frac{p_Z(Z)}{p_{Z}(Y)}\right) + \ln\left(1-\gamma_n\ln(p_Z(Y))p_Z(Y)^{-\theta_n(Y)}\right) \hspace{-0.2em}\bigg)^2\bigg]. 
\end{IEEEeqnarray*}
To prove that the variance of the information density approaches zero, it thus suffices to prove that both expectations on the right-hand side of \eqref{main_computation_variance} approach zero. We start with the second expectation and write it as the sum of two parts:
\begin{IEEEeqnarray}{rCl}
\IEEEeqnarraymulticol{3}{l}{
\mathbb{E}\left[\left(\ln\left(1-\gamma_n\ln(p_Z(Y))p_Z(Y)^{-\theta_n(Y)}\right)\right)^2\right]
}\nonumber\\*
&=&\mathbb{P}\left\{p_Z(Y)\leq1\right\}\nonumber\\
&&\cdot\,\mathbb{E}\left[\left(\ln\left(1-\gamma_n\ln(p_Z(Y))p_Z(Y)^{-\theta_n(Y)}\right)\right)^2\middle|\, p_Z(Y)\le 1 \right]\nonumber\\
&&{}+\mathbb{P}\left\{p_Z(Y)\geq1\right\}\nonumber\\
&&\cdot\,\mathbb{E}\left[\left(\ln\left(1-\gamma_n\ln(p_Z(Y))p_Z(Y)^{-\theta_n(Y)}\right)\right)^2\middle|\, p_Z(Y)\ge 1 \right].\nonumber\\
&& ~\label{eq:E87}
\end{IEEEeqnarray}
For the second part, we notice that, for any $y$ such that $p_Z(y)\geq 1$ (if such $y$ exists), by \eqref{eq:bounded},
\begin{IEEEeqnarray}{c}
\ln(p_Z(y))p_Z(y)^{-\theta_n(y)}
\leq \ln(p_Z(y)) \leq \max\{ 0, \ln(b)\}, \IEEEeqnarraynumspace
\end{IEEEeqnarray}
therefore
\begin{IEEEeqnarray}{rCl}
\IEEEeqnarraymulticol{3}{l}{
\mathbb{E}\left[\left(\ln\left(1-\gamma_n\ln(p_Z(Y))p_Z(Y)^{-\theta_n(Y)}\right)\right)^2\middle|\, p_Z(Y)\ge 1 \right]
}\nonumber\\*~~~~~~~~~~~~~~~~~~~~
\label{used_for_numerator_for_key}
&\leq&\Big(\ln\bigl(1-\gamma_n\max\left\{0,\ln(b)\right\}\bigr)\Big)^2,\IEEEeqnarraynumspace
\end{IEEEeqnarray}
which tends to zero as $n\to\infty$. We now bound the first term on the right-hand side of \eqref{eq:E87} as follows:
\begin{IEEEeqnarray}{rCl}
\IEEEeqnarraymulticol{3}{l}{
\mathbb{P}\left\{ p_Z(Y)\leq1\right\}
} \nonumber\\*
\IEEEeqnarraymulticol{3}{l}{\,\,{}\cdot \mathbb{E}\left[\left(\ln\left(1-\gamma_n\ln(p_Z(Y))p_Z(Y)^{-\theta_n(Y)}\right)\right)^2\middle|\, p_Z(Y)\le 1\right]}\nonumber\\
& \le &
\mathbb{P}\left\{ p_Z(Y)\leq1\right\}\nonumber\\
&&{}\cdot\mathbb{E}\left[\left(\gamma_n\ln(p_Z(Y))p_Z(Y)^{-\theta_n(Y)} \right)^2\middle|\, p_Z(Y)\le 1 \right] \nonumber\\
&\le& \gamma_n^2 \int_{\mathbb{R}} \left(\ln(p_Z(y))\right)^2 p_Z(y)^{-2\theta_n(y)} p_{Y}(y) \dd y \nonumber\\
& = & \gamma_n^2 \alpha_n \int_{\mathbb{R}}  \left(\ln(p_Z(y))\right)^2 p_Z(y)^{1-\gamma_n -2 \theta_n(y)}\dd y. 
\label{eq:pZles1}
\end{IEEEeqnarray}
By Lemma~\ref{lemma_integrability}, for large enough $n$, the integral in the above expression is finite. Since $\alpha_n\to 1$ and $\gamma_n\to 0$ as $n\to\infty$, we conclude that the right-hand side of \eqref{eq:pZles1} tends to zero. We have thus shown that both terms on the right-hand side of \eqref{eq:E87} tend to zero and, therefore, the second expectation on the right-hand side of \eqref{main_computation_variance} tends to zero as $n\to\infty$. 

We now consider the first expectation on the right-hand side of \eqref{main_computation_variance}, which can be written as
\begin{IEEEeqnarray}{rCl}
\IEEEeqnarraymulticol{3}{l}{
\mathbb{E}\left[\left(\ln\left(\frac{p_Z(Z)}{p_Z(Y)}\right)\right)^2\right]
}\nonumber\\*~~~~~~
&=&\mathbb{E}\left[\left(\ln(p_Z(Z))\right)^2\right]-2\,\mathbb{E}\left[\ln(p_Z(Z))\ln(p_Z(Y))\right]\nonumber\\*
&&{}+ \mathbb{E}\left[\left(\ln(p_Z(Y))\right)^2\right]. \label{eq:threeEs}
\end{IEEEeqnarray}
We first consider the third term on the right-hand side:
\begin{IEEEeqnarray}{rCl}
\IEEEeqnarraymulticol{3}{l}{
\mathbb{E}\left[\left(\ln(p_Z(Y))\right)^2\right]
}\nonumber\\*~~~~~~
& = & \int_{\mathbb{R}} p_{Y}(y) \left( \ln (p_Z(y))\right)^2 \dd y\nonumber\\
& = & \int_{\mathbb{R}}\alpha_n p_{Z}(y) (1-\gamma_n \ln(p_Z(y))p_{Z}(y)^{-\theta_n(y)})\nonumber\\*
&&~~~~~~~~~~~~~~~~~~~~~~~~~~~{}\cdot(\ln(p_Z(y)))^2 \dd y. \IEEEeqnarraynumspace
\end{IEEEeqnarray}
By Lemma~\ref{lemma_integrability} and the fact that $\alpha_n\to 1$ as $n\to\infty$, we have
\begin{equation}\label{eq:int92}
    \lim_{n\to\infty} \mathbb{E}\left[\left(\ln(p_Z(Y))\right)^2\right] = \mathbb{E}\left[\left(\ln(p_Z(Z))\right)^2\right].
\end{equation}

We next consider the second term on the right-hand side of \eqref{eq:threeEs}.
We can write this expectation as a double (Riemann-Stieltjes) integral:
\begin{IEEEeqnarray}{rCl}
\IEEEeqnarraymulticol{3}{l}{
\mathbb{E}\left[\ln(p_Z(Z))\ln(p_Z(Y))\right]
}\nonumber\\*
&=& \int_{\mathbb{R}}\int_{\mathbb{R}} p_Z(y-x)  \ln(p_Z(y-x))\ln(p_Z(y))\dd y\, \dd P_X(x).\nonumber\\
&&~
\end{IEEEeqnarray}
Since, by Assumption~\ref{assum}, $t\mapsto p_{Z}(t)\ln(p_Z(t))$ is uniformly continuous on $\textnormal{supp}(p_Z)$, we have that the family of functions $\{t\mapsto p_Z(y-t)\ln(p_Z(y-t))\}_{y}$ is pointwise equicontinuous wherever $y-t\in\textnormal{supp}(p_Z)$. 
Again by Assumption~\ref{assum}, they are also uniformly bounded. By Lemma~\ref{weak_convergence_p_xn} and \cite[Theorem 3.1]{rao1962relations}, \cite[Theorem 8.2.18]{bogachev2007measure}, the following limit holds uniformly over $y\in\mathbb{R}$:
\begin{equation}
\lim_{n\to\infty} \int_{\mathbb{R}} p_Z(y-x) \ln(p_Z(y-x))\dd P_X(x)= p_Z(y)\ln (p_Z(y)).
\end{equation}
Therefore,
\begin{IEEEeqnarray}{rCl}
\IEEEeqnarraymulticol{3}{l}{
\lim_{n\rightarrow\infty}  \int_{\mathbb{R}} \int_{\mathbb{R}}p_{Z}(y-x)\ln(p_Z(y-x))\ln(p_Z(y)) \dd P_X(x)\,\dd y  
}\nonumber\\* 
& = &  \int_{\mathbb{R}} \ln(p_Z(y))  \lim_{n\rightarrow\infty} \int_{\mathbb{R}}p_{Z}(y-x)\ln(p_Z(y-x)) \dd P_X(x)\,\dd y  \nonumber\\
    &=& \int_{\mathbb{R}}\ln(p_Z(y)) p_{Z}(y) \ln (p_Z(y)) \, \dd y \nonumber\\
    \label{first_term}
    &=&\mathbb{E}\left[\left(\ln(p_Z(Z))\right)^2\right]. \label{eq:int101}
\end{IEEEeqnarray}
Combining \eqref{eq:threeEs}, \eqref{eq:int92}, and \eqref{eq:int101}, we obtain
\begin{equation}
\lim_{n\to\infty} \mathbb{E}\left[\left(\ln\left(\frac{p_Z(Z)}{p_Z(Y)}\right)\right)^2\right] = 0.
\end{equation}

We have now shown that both expectations on the right-hand side of \eqref{main_computation_variance} tend to zero as $n\to\infty$. Hence the variance in \eqref{chebyshev} tends to zero as $n\to\infty$, establishing \eqref{lower_boud_L} and completing the proof.
\end{IEEEproof}

\section{Examples}
\label{section:examples}

In this section we apply Theorems~\ref{theorem_converse} and~\ref{theorem_information_spectrum} to specific noise distributions. As before, we consider the channel \eqref{eq:channel} with the noise sequence being i.i.d.

\subsection{Exponential noise}
\label{example_exponential_noise}

Let the noise random variable $Z$ have the exponential distribution of parameter $\Lambda>0$:
\begin{equation}
\label{exponential_noise}
p_Z(z)=\Lambda e^{-\Lambda z},\qquad z\in\mathbb{R}^+.
\end{equation}
This distribution satisfies both \eqref{eq:integrable0}--\eqref{eq:integrable2} and Assumption~\ref{assum}. Verifying \eqref{eq:integrable0}--\eqref{eq:integrable2} and Assumption~\ref{assum} part~\ref{item:ass1}) is straightforward. For part~\ref{item:ass2}), we note that $z\mapsto p_Z(z)\ln(p_Z(z))$ has a bounded derivative, therefore, by the mean value theorem, it is uniformly continuous. To check part~\ref{item:ass3}), 
for any $\gamma>0$, we notice that $\tilde{Z}$ defined in \eqref{def_z_tilde} has the exponential distribution of parameter $(1-\gamma)\Lambda$. It was shown in \cite{exponential_verdu} that there exists $X$ independent of $Z$ such that $\tilde{Z}=X+Z$.

We can hence apply Theorem~\ref{theorem_information_spectrum} to obtain
\begin{IEEEeqnarray*}{rCl}
L 
& = & \sqrt{2} \sqrt{ \textnormal{Var}\left[ \ln(p_Z(Z))\right] }\\
& = & \sqrt{2} \sqrt{ \textnormal{Var}\left[ \ln(\Lambda) - \Lambda Z\right]} \\ 
& = & \sqrt{2} \sqrt{\textnormal{Var} \left[ \Lambda Z \right] }\\
& = & \sqrt{2}. \yesnumber
\end{IEEEeqnarray*}

\subsection{Generalized Gaussian noise}
\label{example_gg_noise}

Consider $Z$ having the generalized Gaussian distribution \cite{nadarajah2005,generalized_gaussian_distributions_short,generalized_gaussian_distributions} of parameters $p>0$ and  $\sigma>0$:
\begin{equation}
\label{generalized_gaussian_noise}
p_Z(z)=\frac{c_p}{\sigma}e^{-\frac{|z|^p}{2\sigma^p}},\qquad z\in\mathbb{R},
\end{equation}
where
\begin{equation}
\label{definition_c_p}
c_p=\frac{p}{2^{\frac{p+1}{p}}\Gamma(\frac{1}{p})},
\end{equation}
with $\Gamma(\cdot)$ denoting the gamma function. The problem of characterizing the scaling constant $L$ for this noise distribution was studied in our previous work \cite{bouette2023covert};
here we shall obtain the same results by directly applying Theorems~\ref{theorem_converse} and~\ref{theorem_information_spectrum}.

First we notice that \eqref{eq:integrable0}--\eqref{eq:integrable2} are satisfied for all $\sigma>0$ and $p>0$. Therefore, we can apply Theorem \ref{theorem_converse} to obtain
\begin{IEEEeqnarray}{rCl}
\label{L_for_gg_noise}
L & \leq & \sqrt{2} \sqrt{\textnormal{Var}\left[\frac{|Z|^p}{2 \sigma^p}\right] }= \sqrt{\frac{2}{p}},
\end{IEEEeqnarray}
recovering \cite[Theorem 2]{bouette2023covert}. 

When $p=2$, \eqref{generalized_gaussian_noise} becomes the Gaussian distribution, which satisfies Assumption~\ref{assum}. Theorem~\ref{theorem_information_spectrum} then implies $L=1$, recovering \cite[Theorem 5]{fundamental_covertness}.

When $p\in(0,1]$, by \cite{generalized_gaussian_distributions} we know that Assumption~\ref{assum} part~\ref{item:ass3}) is satisfied. Indeed, $\tilde{Z}$ defined in \eqref{def_z_tilde} is also generalized Gaussian with the same $p$ but a different $\sigma$; by \cite{generalized_gaussian_distributions}, the generalized Gaussian distribution is self-decomposable when $0<p\le1$, meaning that there exists $X$ independent of $Z$ with $\tilde{Z}=X+Z$. The remaining parts of Assumption~\ref{assum} are straightforward to verify.
We can therefore apply Theorem~\ref{theorem_information_spectrum} to obtain
\begin{equation}
L = \sqrt{\frac{2}{p}},\qquad 0<p\le 1,
\end{equation}
recovering \cite[Theorem 3]{bouette2023covert}.

\subsection{Generalized gamma noise}

Consider $Z$ following the generalized gamma distribution \cite[Appendix A.3]{infinite_divisible}, \cite{stacy1962generalization} of parameters $r,\sigma,\beta>0$:
\begin{equation}
\label{density_generalized_gamma}
    p_{Z}(z)=\frac{\beta}{\Gamma(r)\sigma^{\beta r}}z^{\beta r-1}e^{-\left(\frac{z}{\sigma}\right)^{\beta}},\qquad z\in\mathbb{R}^+
\end{equation}
with $\Gamma(\cdot)$ denoting the gamma function. 

One can easily verify that $p_Z$ satisfies \eqref{eq:integrable0}--\eqref{eq:integrable2}, therefore, by Theorem \ref{theorem_converse}, we have
\begin{equation}
\label{L_for_ggamma_noise}
L\leq\sqrt{2}\sqrt{\left(r-\frac{1}{\beta}\right)^2\psi'(r)-r+\frac{2}{\beta}},
\end{equation}
with $\psi'$ denoting the first derivative of the digamma function~$\psi$:
\begin{equation}
\psi(r) =\frac{\Gamma'(r)}{\Gamma(r)},\quad r\in \mathbb{R}.
\end{equation}
Details of the computation can be found in Appendix~\ref{computation_L_ggamma}.

\begin{remark}
When $\beta=1$, \eqref{density_generalized_gamma} reduces to the gamma distribution. When $r=\frac{1}{\beta}$, \eqref{density_generalized_gamma} reduces to the generalized Gaussian distribution~\eqref{generalized_gaussian_noise} with $p=\beta$, but restricted to $\mathbb{R}^+$, and \eqref{L_for_ggamma_noise} becomes the same as \eqref{L_for_gg_noise}.
We cannot apply Theorem~\ref{theorem_information_spectrum} to obtain an exact characterization of $L$ because the generalized gamma distribution is not known to satisfy Assumption \ref{assum}, in particular its part~\ref{item:ass3}).
\end{remark}

\section{Bounds on the Key Length}
\label{section:key_length}

In Section \ref{section:lower_bound_L}, we assumed that an arbitrarily long key was shared between the transmitter and the receiver. In this section, we show that a finite key is sufficient to achieve the optimal $L$ given in Theorem \ref{theorem_information_spectrum}. We provide an upper bound on the required key length in terms of $n$ and refine this upper bound when the noise has Gaussian or exponential distribution. 

\begin{proposition}
\label{proposition:minimum_key_length_general_case}
For the memoryless additive-noise channel \eqref{eq:channel}, if $p_Z$ satisfies \eqref{eq:integrable0}--\eqref{eq:integrable2} as well as Assumption \ref{assum}, 
then there exists a sequence of codes that asymptotically achieves the optimal scaling
factor $L$ of Theorem \ref{theorem_information_spectrum} with key lengths satisfying
\begin{IEEEeqnarray}{rCl} \label{eq:keyOn}
    \ln\left|\mathcal{K}\right|=O(n).
\end{IEEEeqnarray}
\end{proposition}

\begin{proposition}
\label{proposition:minimum_key_length_gaussian_exponential}
For $P_Z$ being a Gaussian or exponential distribution, 
\eqref{eq:keyOn} can be strengthened to
\begin{IEEEeqnarray}{rCl}
    \ln\left|\mathcal{K}\right|=o(\sqrt{n}).
\end{IEEEeqnarray}
\end{proposition}

The proofs of Propositions \ref{proposition:minimum_key_length_general_case} and \ref{proposition:minimum_key_length_gaussian_exponential} can be found in Appendix \ref{appendix_gg_minimum_length_key}.

\begin{remark} 
For Gaussian noise, the result of Proposition~\ref{proposition:minimum_key_length_gaussian_exponential} is essentially known, albeit in settings with several technical differences from ours; for example, apply \cite[Theorem 6]{bloch_resolvability} to the special case where (in the notation therein) $P_0=Q_0$ and $P_1=Q_1$. If the eavesdropper's channel is noisier than the receiver's channel, then one can show that no key is needed. This would be similar to \cite[Theorem 6]{bloch_resolvability} with ${D}(Q_1\|Q_0) < {D}(P_1\|P_0)$. Some other related results on the key length in Gaussian covert communication can be found in \cite{zhang2019undetectable, bloch2020undetectable,wang2021covert}.
\end{remark}

\section{Concluding Remarks}
\label{section:conclusion}

There are not many examples of additive-noise channels whose capacity under a certain input cost constraint \cite{gallager87,verdu90} admits a closed-form expression, notably the AWGN channel with a second-moment constraint, the exponential-noise channel with a first-moment constraint~\cite{exponential_verdu}, the channel with Cauchy noise and logarithmic constraint \cite{fahs2014cauchy}, and some generalized Gaussian-noise channels \cite{generalized_gaussian_distributions_short}. Some works provide bounds on the capacity or show properties of optimal input distributions, e.g., \cite{de2017capacity,fahs_properties_nodate}.

In contrast, we are able to derive a simple expression for the scaling constant $L$ for covert communication over rather general additive-noise channels. This is because, in a sense, the covertness condition~\eqref{covert_communication_hypothesis} translates to a constraint on the output that is naturally ``fitted'' to the noise distribution,\footnote{For example, for Gaussian noise, it translates into a second-moment constraint, and for exponential noise into a first-moment constraint.} which in turn allows for a precise characterization of (or an elegant upper bound on) $L$.

It is not clear to us how to generalize Theorems~\ref{theorem_converse} and~\ref{theorem_information_spectrum} to scenarios where the receiver and the eavesdropper face different noise distributions, because the cost constraint would now be fitted to the eavesdropper's noise, not the receiver's.
A special case where such generalization is straightforward is the Gaussian wiretap channel: the receiver and the eavesdropper are corrupted by Gaussian noise with different variances $\sigma^2$ and $\sigma_e^2$, respectively. In this case, it is well known that $L=\sigma_e^2/\sigma^2$.

Validity of the formula \eqref{eq:Lequal} is limited by Assumption~\ref{assum}, notably its part \ref{item:ass3}). For example,  \ref{item:ass3}) is not known to be true for some generalized Gaussian distributions~\cite{generalized_gaussian_distributions}. However, \ref{item:ass3}) being false (or not known to be true or false) for a certain noise PDF does not necessarily imply that the formula \eqref{eq:Lequal} cannot hold for such noise. This is because the input distribution to achieve $L$ need not be unique, i.e., we do not need the PDF of $X+Z$ to be exactly \eqref{def_z_tilde}. For example, for the AWGN channel, \eqref{def_z_tilde} is a Gaussian distribution, and $X$ should also be Gaussian in order to induce this output distribution. However, choosing $X$ to take only two values, $\pm a$ for some $a$ approaching zero as $n\to\infty$, can also attain $L$ \cite{wang19}. If one can show that similar input distributions can attain $L$ on other additive-noise channels, in particular, those that do not satisfy Assumption~\ref{assum}, then one may be able to further extend the validity of \eqref{eq:Lequal}.

\appendices

\section{Uniform Noise}
\label{appendix_uniform_case}

Consider the noise \eqref{eq:uniform}.
Suppose there exists an input distribution that yields an output distribution $P_Y$ such that
\begin{equation}\label{eq:Dfinite}
    {D}(P_{Y}\|P_{Z})<\infty.
\end{equation}
For this to hold, $P_{Y}$ needs to be absolutely continuous with respect to $P_Z$, i.e., $P_{Y}\ll P_Z$, and we have
\begin{IEEEeqnarray}{rCl}
    {D}(P_{Y}\|P_Z)&=&-h(Y)-\int_{\mathcal{S}}p_{Y}(y)\ln(p_Z(y))dy\nonumber\\
    &=&-h(Y)-\ln\left(\frac{1}{\lambda\left(\mathcal{S}\right)}\right)\nonumber\\
    &=&h(Z)-h(Y).\label{eq:DisHZY}
\end{IEEEeqnarray}
On the other hand, since $Y=X+Z$, we must have
\begin{IEEEeqnarray}{rCl}
    I(X;Y)&=&h(Y)- h(Z)\nonumber\\
    &=&-{D}(P_{Y}\|P_Z)
\end{IEEEeqnarray}
where the last step follows by \eqref{eq:DisHZY}. Since $I(X;Y)\ge0$, the only way this can happen is to have $P_Z=P_Y$. That is, the only way to satisfy \eqref{eq:Dfinite} is to have $X=0$ with probability one. It thus follows that covert communication is not possible and therefore, by definition, $L=0$.

\section{Proofs of Propositions \ref{proposition:minimum_key_length_general_case} and \ref{proposition:minimum_key_length_gaussian_exponential}}
\label{appendix_gg_minimum_length_key}

We consider the same random code construction as in the proof of Theorem \ref{proof_theorem_2} in Section \ref{section:lower_bound_L}, where the codewords are generated i.i.d. according to $P_X$ such that $X_n\sim P_X$ satisfies \eqref{eq:ZnXnZ}. (As before, to avoid cumbersome notation, we do not explicitly write the distributions of $X$ and $Y$ as functions of $n$.)
We denote by $\mathsf{C}$ the random codebook and by $P_{Y^n|\mathsf{C}}$ the output distribution conditional on $\mathsf{C}$. Recalling \eqref{eq:covertness2}, we shall find sufficient conditions on $\ln|\mathcal{K}|$ such that 
\begin{equation}\label{eq:ECD0}
\lim_{n\to\infty} \mathbb{E}_{\mathsf{C}}\left[{D}\left(P_{Y^n|\mathsf{C}}\|P_{Y^n}\right)\right]=0,
\end{equation}
which will ensure that $\mathsf{C}$ has the desired covertness property with high probability. This will further guarantee the existence of a good deterministic code $\mathcal{C}$ that is covert and has a small error probability, because we have already shown that, with high probability, the random code $\mathsf{C}$ has an average error probability that tends to zero.

To establish \eqref{eq:ECD0}, we apply the channel resolvability bound of \cite[Theorem 14]{hayashi2016secure}, which asserts that, for $\rho\in(0,1]$, 
\begin{IEEEeqnarray}{rCl}
\IEEEeqnarraymulticol{3}{l}{
\mathbb{E}_{\mathsf{C}}\left[{D}\left(P_{Y^n|\mathsf{C}}\|P_{Y^n}\right)\right]
}\nonumber\\*~~~~~
\label{use_hayashi}
&\leq& \frac{1}{\rho} \ln\left(1+e^{-\rho\ln\left(\left|\mathcal{K}\right|\times\left|\mathcal{M}\right|\right)+ n \Psi(\rho|P_{Y|X},P_{X})}\right) 
\IEEEeqnarraynumspace
\end{IEEEeqnarray}
where 
\begin{IEEEeqnarray}{rCl}
\label{definition_psi}
\Psi(\rho|P_{Y|X},P_{X})&=&\ln \left( \mathbb{E} \left[\left(\frac{p_{Y|X}(Y|X)}{p_{Y}(Y)}\right)^\rho \right]\right).\IEEEeqnarraynumspace
\end{IEEEeqnarray}
(A similar resolvability technique based on \cite[Theorem 14]{hayashi2016secure} was proposed in \cite[Lemma 13]{bloch2020undetectable}.)

\begin{IEEEproof}[Proof of Proposition \ref{proposition:minimum_key_length_general_case}]
For sufficiently small $\rho$, the expectation in \eqref{definition_psi} can be upper-bounded as
\begin{IEEEeqnarray}{rCl}
\IEEEeqnarraymulticol{3}{l}{
\mathbb{E} \left[\left(\frac{p_{Y|X}(Y|X)}{p_{Y}(Y)}\right)^\rho \right] 
}\nonumber\\* ~~
\label{use_of_bounded}
&=& \mathbb{E} \left[\left(\frac{p_{Z}(Y-X)}{p_{Y}(Y)}\right)^\rho \right]
\nonumber\\
&\leq& \mathbb{E} \left[\left(\frac{b}{p_{Y}(Y)}\right)^\rho \right] \\
&=& b^{\rho}\int_{\mathbb{R}}p_{Y}(y)^{1-\rho} \dd y \nonumber\\
&=& b^{\rho}\int_{\mathbb{R}}\alpha_n^{1-{\rho}} p_{Z}(y)^{1-\gamma_n(1-\rho)-\rho} \dd y \nonumber\\*
\label{upper_bound_denominator_use_dl}
&=& \alpha_n^{1-{\rho}}b^{\rho}\int_{\mathbb{R}}p_{Z}(y)^{1-\rho}\nonumber\\*
&&\quad{}\cdot\left(1-\gamma_n(1-\rho)\ln(p_Z(y))p_{Z}(y)^{-\theta_n(y)}\right) \dd y \IEEEeqnarraynumspace\\ 
\label{upper_bound_denominator}
&=& b^{\rho} \int_{\mathbb{R}}p_{Z}(y)^{1-\rho} \dd y +O\left(\frac{1}{\sqrt{n}}\right)
\end{IEEEeqnarray}
where \eqref{use_of_bounded} follows by \eqref{eq:bounded}; \eqref{upper_bound_denominator_use_dl} by the Taylor expansion of $p_{Y}(y)$ with the Lagrange form of the remainder \eqref{other_dl}; and \eqref{upper_bound_denominator} by \eqref{eq:alpha22},  \eqref{eq:gamman}, and Lemma \ref{lemma_integrability}. 
For sufficiently small $\rho$, Lemma \ref{lemma_integrability} ensures that the integral in \eqref{upper_bound_denominator} is finite, which in turn implies that \eqref{definition_psi} is bounded. Therefore choosing $\ln\left|\mathcal{K}\right|=O(n)$ ensures that the right-hand side of \eqref{use_hayashi} goes to $0$ as $n\to\infty$, establishing~\eqref{eq:ECD0}.
\end{IEEEproof}

\begin{IEEEproof}[Proof of Proposition \ref{proposition:minimum_key_length_gaussian_exponential}]

\paragraph{Gaussian noise}

Let $Z\sim \mathcal{N}(0,\sigma^2)$ with $\sigma>0$. It is easy to check that the input and output distributions are $X\sim \mathcal{N}\left(0,\frac{\sigma^2 \gamma_n}{1-\gamma_n}\right)$ and $Y\sim \mathcal{N}\left(0,\frac{\sigma^2}{1-\gamma_n}\right)$, respectively, and
\begin{IEEEeqnarray}{rCl}
\IEEEeqnarraymulticol{3}{l}{
\Psi(\rho|P_{Y|X},P_{X})
}\nonumber\\* ~~
&=&\ln\left(\int_{\mathbb{R}}\int_{\mathbb{R}}p_{X,Y}(x,y)\left(\frac{p_{Y|X}(y|x)}{p_{Y}(y)}\right)^\rho \dd x \, \dd y\right)\nonumber\\
&=&\ln\left(\int_{\mathbb{R}}\int_{\mathbb{R}}p_{Z}(y-x)p_{X}(x)\left(\frac{p_{Z}(y-x)}{p_{Y}(y)}\right)^\rho \dd x \, \dd y\right)\nonumber\\
&=&\ln\left(\sizecorr{\frac{e^{-\frac{(y-x)^2}{2\sigma^2}}}{e^{-\frac{y^2}{2\sigma^2\frac{1}{1-\gamma_n}}}}}\int_{\mathbb{R}}\int_{\mathbb{R}}\left(\frac{1}{\sqrt{2\pi}\sigma}\right)^2 \sqrt{\frac{1-\gamma_n}{\gamma_n}}e^{-\frac{(y-x)^2}{2\sigma^2}}e^{-\frac{x^2}{2\sigma^2\frac{\gamma_n}{1-\gamma_n}}}\right.\nonumber\\*
&&\qquad\qquad\qquad \left.\cdot \left(\frac{1}{\sqrt{1-\gamma_n}}\frac{e^{-\frac{(y-x)^2}{2\sigma^2}}}{e^{-\frac{y^2}{2\sigma^2\frac{1}{1-\gamma_n}}}}\right)^\rho \dd x \, \dd y\right)\nonumber\\
\label{interchangement_order_integrals_gaussian}
&=&\ln\left(\int_{\mathbb{R}}\int_{\mathbb{R}}\left(\frac{1}{\sqrt{2\pi}\sigma}\right)^2 \frac{(1-\gamma_n)^{\frac{1}{2}-\frac{\rho}{2}}}{\sqrt{\gamma_n}} e^{-\frac{\left(y-\frac{1+\rho}{1+\rho\gamma_n}x\right)^2}{2\sigma^2\frac{1}{1+\rho\gamma_n}}}\right.\nonumber\\*
&&\qquad\qquad\qquad\left.\sizecorr{e^{-\frac{\left(y-\frac{1+\rho}{1+\rho\gamma_n}x\right)^2}{2\sigma^2\frac{1}{1+\rho\gamma_n}}}}\cdot e^{-\frac{x^2}{2\sigma^2}\frac{(1-\gamma_n)(1-\rho^2\gamma_n)}{\gamma_n(1+\rho\gamma_n)}} \dd y \, \dd x\right)\\
&=&\ln\left(\frac{(1-\gamma_n)^{\frac{1}{2}-\frac{\rho}{2}}}{\sqrt{\gamma_n}} \frac{1}{\sqrt{1+\rho\gamma_n}} \sqrt{\frac{\gamma_n(1+\rho\gamma_n)}{(1-\gamma_n)(1-\rho^2\gamma_n)}} \right)\nonumber\\
\label{step_gallager_function_gaussian}
&=&\ln\left(\frac{(1-\gamma_n)^{-\frac{\rho}{2}}}{\sqrt{1-\rho^2\gamma_n}}\right), 
\end{IEEEeqnarray}
where \eqref{interchangement_order_integrals_gaussian} follows by Fubini's theorem.

If the message rate scales according to the optimal scaling constant $L$ as in Theorem \ref{theorem_information_spectrum}, i.e., $\lim\limits_{n\rightarrow \infty} \frac{\ln\left|\mathcal{M}\right|}{\sqrt{\Delta}\sqrt{n}}=1$, then there exists a positive sequence $\{\xi_n\}$ such that $\xi_n=o(\sqrt{n})$ and $\ln\left|\mathcal{M}\right|\geq\sqrt{\Delta}\sqrt{n}-\xi_n$. 
Let $\{\rho_n\}, \rho_n\in \left(0,1\right)$ for any $n$, be such that $\rho_n \rightarrow 0$ and $\rho_n\xi_n\rightarrow \infty$ when $n\rightarrow \infty$. Continuing from \eqref{step_gallager_function_gaussian}, we obtain
\begin{IEEEeqnarray}{rCl}
\IEEEeqnarraymulticol{3}{l}{\Psi(\rho_n|P_{Y|X},P_{X})}\nonumber\\* \qquad
&=&-\frac{\rho_n}{2} \ln\left(1-\gamma_n\right)-\frac{1}{2}\ln\left(1-\rho_n^2\gamma_n\right)\nonumber\\
&=&-\frac{\rho_n}{2} \left(-\gamma_n+O(\gamma_n^2)\right) - \frac{1}{2}\left(-\rho_n^2\gamma_n+O(\rho_n^4\gamma_n^2)\right)\nonumber\\
&=&\left(\frac{\rho_n}{2}+\frac{\rho_n^2}{2}\right)\gamma_n+O\left(\rho_n\gamma_n^2\right)\nonumber\\
\label{end_gallager_function_gaussian}
&=&\rho_n\left(1+\rho_n\right)\sqrt{\frac{\Delta}{n}}+O\left(\frac{\rho_n}{n}\right),
\end{IEEEeqnarray}
where \eqref{end_gallager_function_gaussian} follows by recalling the expression of $\gamma_n$ in \eqref{eq:gamman}.
We have now established the upper bound
\begin{IEEEeqnarray}{rCl}
\IEEEeqnarraymulticol{3}{l}{
\mathbb{E}_{\mathsf{C}}\left[{D}\left(P_{Y^{n}|\mathsf{C}}||P_{Y^{n}}\right)\right]
}\nonumber\\* ~~~~~
    &\leq& \frac{1}{\rho_n} \ln\left(1+ e^{-\rho_n( \ln\left|\mathcal{K}\right|+\ln\left|\mathcal{M}\right|) + \rho_n(1+\rho_n) \sqrt{\Delta}\sqrt{n} +O (\rho_n)}\right)\nonumber\\
    &\leq& \frac{1}{\rho_n} e^{-\rho_n\left( \ln\left|\mathcal{K}\right|-\rho_n \sqrt{\Delta}\sqrt{n} -\xi_n +O(1)\right)}.
\end{IEEEeqnarray}
A key length
$\ln\left|\mathcal{K}\right|=\rho_n\sqrt{\Delta}\sqrt{n}+2\xi_n$ would be sufficient to ensure \eqref{eq:ECD0}.
Since $\rho_n\rightarrow 0$ and $\xi_n = o\left(\sqrt{n}\right)$, it follows that $\ln\left|\mathcal{K}\right|=o\left(\sqrt{n}\right)$.

\paragraph{Exponential noise}

For $Z$ having the exponential distribution \eqref{exponential_noise} of parameter $\Lambda>0$, which we denote $\mathcal{E}(\Lambda)$, the target output distribution is $Y\sim \mathcal{E}((1-\gamma_n)\Lambda)$. The input distribution that induces this output distribution is a mixture of a point mass at $0$ and an exponential distribution \cite{exponential_verdu}:
\begin{IEEEeqnarray}{rCl}
P_X(x) = \begin{cases} 0, & x<0\\ 1- \gamma_n e^{-(1-\gamma_n)\Lambda x}, & x\ge 0.\end{cases} \IEEEeqnarraynumspace
\end{IEEEeqnarray}
We compute $\Psi(\rho|P_{Y|X},P_{X})$ as follows:
\begin{IEEEeqnarray}{rCl}
\IEEEeqnarraymulticol{3}{l}{
\Psi(\rho|P_{Y|X},P_{X})
}\nonumber\\*
&=&\ln\left(\int_{\mathbb{R}}\int_{\mathbb{R}} p_{Z}(y-x)\left(\frac{p_{Z}(y-x)}{p_{Y}(y)}\right)^\rho \dd P_{X}( x)\, \dd y\right)\nonumber\\
&=&\ln\Bigg((1-\gamma_n) \int_{0}^\infty \Lambda  e^{-\Lambda y}  \left(\frac{1}{1-\gamma_n}\frac{e^{-\Lambda y}}{e^{-\Lambda (1-\gamma_n)y}}\right)^\rho \dd y \nonumber\\*
&&\quad\quad{}+ \gamma_n\int_{0}^\infty\int_{0}^y\Lambda^2 (1-\gamma_n) e^{-\Lambda (y-x)} e^{-\Lambda (1-\gamma_n) x} \nonumber\\*
&&\qquad\qquad\qquad\qquad\quad\quad{}\cdot\bigg(\frac{1}{1-\gamma_n}\frac{e^{-\Lambda (y-x)}}{e^{-\Lambda (1-\gamma_n)y}}\bigg)^\rho \dd x \, \dd y\Bigg)\nonumber\\
&=&\ln\Bigg((1-\gamma_n)^{1-\rho} \int_{0}^\infty \Lambda  e^{-\Lambda (1+\rho\gamma_n) y} \dd y \nonumber\\
\IEEEeqnarraymulticol{3}{r}{{}+ \gamma_n\int_{0}^\infty\int_{0}^y\Lambda^2 (1-\gamma_n)^{1-\rho} e^{-\Lambda (1+\gamma_n \rho)y} e^{\Lambda (\gamma_n+\rho) x} \dd x \, \dd y\Bigg)}\nonumber\\
&=&\ln\Bigg( \frac{(1-\gamma_n)^{1-\rho}}{1+\rho\gamma_n}\nonumber\\
\IEEEeqnarraymulticol{3}{r}{{} + \gamma_n\int_{0}^\infty\Lambda \frac{(1-\gamma_n)^{1-\rho}}{\gamma_n+\rho} e^{-\Lambda (1+\gamma_n \rho)y} \left(e^{\Lambda (\gamma_n+\rho) y} - 1\right) \dd y \Bigg)}\nonumber\\
&=&\ln\Bigg(1-\gamma_n + \gamma_n \frac{(1-\gamma_n)^{1-\rho}}{\gamma_n+\rho} \nonumber\\
&&~~~~~~{}\cdot\left(\frac{1}{1+\gamma_n\rho-\gamma_n-\rho}-\frac{1}{1+\gamma_n\rho}\right) + o(\gamma_n)\Bigg)\nonumber\\
&=& \frac{\rho}{1-\rho} \gamma_n + o(\gamma_n)\nonumber\\
\label{end_gallager_function_exponential}
&=& \frac{\sqrt{2}\rho}{1-\rho}\sqrt{\frac{\Delta}{n}} +o\left(\frac{1}{\sqrt{n}}\right),
\end{IEEEeqnarray}
where \eqref{end_gallager_function_exponential} follows by recalling the expression of $\gamma_n$ in \eqref{eq:gamman}. 
By a similar reasoning to the Gaussian case, it follows that $\ln\left|\mathcal{K}\right|=o\left(\sqrt{n}\right)$ suffices to ensure \eqref{eq:ECD0}.
\end{IEEEproof}

\section{Derivation of \eqref{L_for_ggamma_noise}}
\label{computation_L_ggamma}

By Theorem 1, 
\begin{IEEEeqnarray}{rCl} \label{eq:122}
L &\leq& \sqrt{2}\sqrt{\textnormal{Var}[\ln(p_Z(Z))]} \nonumber\\
&=& \sqrt{2}\sqrt{\mathbb{E}[\ln(p_Z(Z))^2] - \big(h(Z)\big)^2}, 
\end{IEEEeqnarray}
where
\begin{IEEEeqnarray}{rCl}
\label{entropy_squared_gamma}
    h(Z)&=&\ln\left(\frac{\Gamma(r)\sigma}{\beta}\right)+r+\left(\frac{1}{\beta}-r\right)\psi(r).
\end{IEEEeqnarray}
The other term on the right-hand side of \eqref{eq:122} can be written as
\begin{IEEEeqnarray}{rCl}
\IEEEeqnarraymulticol{3}{l}{
\mathbb{E}[\ln(p_Z(Z))^2]
}\nonumber\\*
&=&\int_{0}^{\infty}p_Z(z)\nonumber\\*
&&\qquad{}\cdot \left(\ln\left(\frac{\beta}{\Gamma(r)\sigma^{\beta r}}\right)+(\beta r-1)\ln(z)-\left(\frac{z}{\sigma}\right)^{\beta}\right)^2\dd z\nonumber\\
&=&\left(\ln\left(\frac{\beta}{\Gamma(r)\sigma^{\beta r}}\right)\right)^2\nonumber\\
\IEEEeqnarraymulticol{3}{r}{\,\,\,\,{}+ \int_{0}^{\infty}p_Z(z)\Bigg(2\ln\left(\frac{\beta}{\Gamma(r)\sigma^{\beta r}}\right)\left((\beta r -1)\ln(z)-\left(\frac{z}{\sigma}\right)^{\beta}\right)} \nonumber\\
&&\qquad\qquad\qquad{}+(\beta r-1)^2\ln(z)^2\nonumber\\
&&\qquad\qquad\qquad\quad{}-2 (\beta r -1)\ln(z)\left(\frac{z}{\sigma}\right)^{\beta}+\left(\frac{z}{\sigma}\right)^{2\beta}\Bigg)\dd z\nonumber\\
&=& \left(\ln\left(\frac{\beta}{\Gamma(r)\sigma^{\beta r}}\right)\right)^2 \nonumber\\
&&{}+2\ln\left(\frac{\beta}{\Gamma(r)\sigma^{\beta r}}\right)\left((\beta r-1)\mathbb{E}\left[\ln(Z)\right]-\mathbb{E}\left[\left(\frac{Z}{\sigma}\right)^\beta\right]\right)\nonumber\\
\label{computation_ex_ln_squared_generalized_gamma}
&&{}+(\beta r -1)^2\mathbb{E}\left[\ln(Z)^2\right]-2(\beta r-1)\mathbb{E}\left[\ln(Z)\left(\frac{Z}{\sigma}\right)^\beta\right]\nonumber\\
&&{}+\mathbb{E}\left[\left(\frac{Z}{\sigma}\right)^{2\beta}\right].
\end{IEEEeqnarray}
We now proceed to compute all the terms in \eqref{computation_ex_ln_squared_generalized_gamma}:
\begin{IEEEeqnarray}{rCl}
    \mathbb{E}\left[\left(\frac{Z}{\sigma}\right)^{\beta}\right]&=& r, 
\end{IEEEeqnarray}
\begin{IEEEeqnarray}{rCl}
    \mathbb{E}\left[\left(\frac{Z}{\sigma}\right)^{2\beta}\right]&=& (r+1)r,  
\end{IEEEeqnarray}
\begin{IEEEeqnarray}{rCl}
    \mathbb{E}[\ln(Z)]&=& \int_{0}^{\infty}\frac{\beta}{\Gamma(r)\sigma^{\beta r}}z^{\beta r-1}e^{-\left(\frac{z}{\sigma}\right)^{\beta}}\ln(z)\dd z\nonumber\\
    &=& \frac{1}{\sigma}\int_{0}^{\infty}\frac{\beta}{\Gamma(r)}\left(\frac{z}{\sigma}\right)^{\beta r-1}e^{-\left(\frac{z}{\sigma}\right)^{\beta}}
    \ln\left(\frac{z}{\sigma}\right)\dd z+\ln(\sigma)\nonumber\\
    &=&\frac{1}{\beta\Gamma(r)} \int_{0}^{\infty}z^{r-1}e^{-z}\ln(z)\dd z+\ln(\sigma)\nonumber\\
    &=&\frac{\Gamma'(r)}{\beta\Gamma(r)}+\ln(\sigma)\nonumber\\
    &=&\frac{1}{\beta}\psi(r)+\ln(\sigma), \label{digamma}
\end{IEEEeqnarray}
\begin{IEEEeqnarray}{rCl}
    \mathbb{E}[\ln(Z)^2]&=& \int_{0}^{\infty}\frac{\beta}{\Gamma(r)\sigma^{\beta r}}z^{\beta r-1}e^{-\left(\frac{z}{\sigma}\right)^{\beta}}\ln(z)^2 \dd z\nonumber\\
    &=& \frac{1}{\sigma}\int_{0}^{\infty}\frac{\beta}{\Gamma(r)}\left(\frac{z}{\sigma}\right)^{\beta r-1}e^{-\left(\frac{z}{\sigma}\right)^{\beta}}\ln\left(\frac{z}{\sigma}\right)^2\dd z\nonumber\\
    &&{}+2\ln(\sigma)\mathbb{E}[\ln(Z)]-\ln(\sigma)^2\nonumber\\
    &=&\frac{1}{\beta^2\Gamma(r)} \int_{0}^{\infty}z^{r-1}e^{-z}\ln(z)^2\dd z\nonumber\\
    &&{}+2\ln(\sigma)\mathbb{E}[\ln(Z)]-\ln(\sigma)^2\nonumber\\
    &=&\frac{\Gamma''(r)}{\beta^2\Gamma(r)}+2\ln(\sigma)\mathbb{E}[\ln(Z)]-\ln(\sigma)^2\nonumber\\
    &=&\frac{1}{\beta^2}\psi'(r)+\frac{1}{\beta^2}\psi(r)^2+2\frac{\ln(\sigma)}{\beta}\psi(r)+\ln(\sigma)^2,\nonumber\\*
\end{IEEEeqnarray}
\begin{IEEEeqnarray}{rCl}
\IEEEeqnarraymulticol{3}{l}{\mathbb{E}\left[\ln(Z)\left(\frac{Z}{\sigma}\right)^{\beta}\right]}\nonumber\\*
\qquad&=&\int_{0}^{\infty}\frac{\beta}{\Gamma(r)\sigma^{\beta r}}z^{\beta r-1}e^{-\left(\frac{z}{\sigma}\right)^{\beta}}\ln(z)\left(\frac{z}{\sigma}\right)^\beta \dd z\nonumber\\
&=&\frac{1}{\sigma}\int_{0}^{\infty}\frac{\beta}{\Gamma(r)}\left(\frac{z}{\sigma}\right)^{\beta r-1}e^{-\left(\frac{z}{\sigma}\right)^{\beta}}\ln\left(\frac{z}{\sigma}\right)\left(\frac{z}{\sigma}\right)^\beta \dd z\nonumber\\
&&{}+\ln(\sigma)\mathbb{E}\left[\left(\frac{Z}{\sigma}\right)^\beta\right]\nonumber\\
&=&\frac{1}{\beta\Gamma(r)}\int_{0}^{\infty}z^{r}e^{-z}\ln(z)\dd z+r\ln(\sigma)\nonumber\\
&=&\frac{\Gamma'(r+1)}{\beta\Gamma(r)}+r\ln(\sigma)\nonumber\\
&=&\frac{1}{\beta}r\frac{\Gamma'(r)}{\Gamma(r)}+\frac{1}{\beta}+r\ln(\sigma)\nonumber\\
& =& \frac{r}{\beta} \psi(r) +\frac{1}{\beta}+r\ln(\sigma).
\end{IEEEeqnarray}
Combining all numbered equations above yields \eqref{L_for_ggamma_noise}.

\section*{Acknowledgment}

The authors would like to thank Malcolm Egan and Matthieu Bloch for helpful discussions, as well as the anonymous reviewers for their suggestions and comments.

\bibliographystyle{IEEEtran}
\bibliography{IEEEabrv,biblio}

\end{document}